\documentclass[twocolumn]{aastex63}
\usepackage{epsfig}
\usepackage{natbib}
\usepackage{color}
\usepackage{amsmath}
\usepackage[squaren]{SIunits}

\submitjournal{Astronomical Journal}%5/1/2018

\newcommand\gaia{{\em Gaia }}

\newcommand\rgc{$R_{GC}$ }

\shortauthors{Donor et al.}
\shorttitle{OCCAM: IV. Cluster Abundances with APOGEE DR16}

\begin{document}

\title{The Open Cluster Chemical Abundances and Mapping Survey: IV.\\ 
   Abundances for 128 Open Clusters using SDSS/APOGEE DR16}

\correspondingauthor{John Donor}
\email{j.donor@tcu.edu}

\author{John Donor}
\affiliation{Department of Physics \& Astronomy, Texas Christian University,
TCU Box 298840, \\Fort Worth, TX 76129, USA (j.donor, p.frinchaboy@tcu.edu)}
%%%%%%%%%%%%%%%%%%%%%%%%%%%%%%%%%%%%%%%%%%%%%%%%%%%%%%%%%%%%
\author[0000-0002-0740-8346]{Peter M. Frinchaboy}
\affiliation{Department of Physics \& Astronomy, Texas Christian University,
TCU Box 298840, \\Fort Worth, TX 76129, USA (j.donor, p.frinchaboy@tcu.edu)}
%%%%%%%%%%%%%%%%%%%%%%%%%%%%%%%%%%%%%%%%%%%%%%%%%%%%%%%%%%%%
\author{Katia Cunha}
\affil{Steward Observatory, The University of Arizona, 933 North Cherry Avenue, Tucson, AZ 85721-0065, USA}
\affil{Observat\'orio Nacional, Rua General Jos\'e Cristino, 77, 20921-400 S\~ao Crist\'ov\~ao, Rio de Janeiro, RJ, Brazil}

% %%%%%%%%%%%%%%%%%%%%%%%%%%%%%%%%%%%%%%%%%%%%%%%%%%%%%%%%%%%%
\author[0000-0003-2321-950X]{Julia E. O\rq{}Connell}
\affiliation{Departamento de Astronom{\'{\i}}a, Universidad de Concepci{\'o}n, Casilla 160-C, Concepci{\'o}n, Chile}

% %%%%%%%%%%%%%%%%%%%%%%%%%%%%%%%%%%%%%%%%%%%%%%%%%%%%%%%%%%%%
\author[0000-0002-0084-572X]{Carlos Allende Prieto}
\affiliation{Instituto de Astrof{\'i}sica de Canarias, V{\'i}a L\'actea S/N, 38205 La Laguna, Tenerife, Spain}
\affiliation{Universidad de La Laguna, Departamento de Astrof{\'i}sica, 30206 La Laguna, Tenerife, Spain}

% %%%%%%%%%%%%%%%%%%%%%%%%%%%%%%%%%%%%%%%%%%%%%%%%%%%%%%%%%%%%
\author{Andr{\'e}s Almeida}
\affiliation{Instituto de Investigaci{\'o}n Multidisciplinario en Ciencia y Tecnolog{\'i}a,\\ Universidad de La Serena, Benavente 980, La Serena, Chile} %(aalmeida@userena.cl)}

% %%%%%%%%%%%%%%%%%%%%%%%%%%%%%%%%%%%%%%%%%%%%%%%%%%%%%%%%%%%%
\author{Friedrich Anders}
\affiliation{Departament de F\'isica Qu\`antica i Astrof\'isica, Universitat de Barcelona, IEEC-UB, Mart\'i i Franqu\`es 1 08028 Barcelona, Spain}

% %%%%%%%%%%%%%%%%%%%%%%%%%%%%%%%%%%%%%%%%%%%%%%%%%%%%%%%%%%%%
\author[0000-0002-1691-8217]{Rachael Beaton}
\altaffiliation{Hubble Fellow}
\altaffiliation{Carnegie-Princeton Fellow}
\affiliation{Department of Astrophysical Sciences, Princeton University, 4 Ivy Lane, Princeton, NJ~08544}
\affiliation{The Observatories of the Carnegie Institution for Science, 813 Santa Barbara St., Pasadena, CA~91101}

% %%%%%%%%%%%%%%%%%%%%%%%%%%%%%%%%%%%%%%%%%%%%%%%%%%%%%%%%%%%%

\author[0000-0002-3601-133X]{Dmitry Bizyaev}
\affiliation{Apache Point Observatory and New Mexico State
    University, P.O. Box 59, Sunspot, NM, 88349-0059, USA} %(kpan, dmbiz@apo.nmsu.edu)}
\affiliation{Sternberg Astronomical Institute, Moscow State
    University, Moscow, Russia}

% %%%%%%%%%%%%%%%%%%%%%%%%%%%%%%%%%%%%%%%%%%%%%%%%%%%%%%%%%%%%
\author[0000-0002-8725-1069]{Joel R. Brownstein}
\affiliation{Department of Physics \& Astronomy, University of Utah, 115 S. 1400 E., Salt Lake City, UT 84112, USA}

% %%%%%%%%%%%%%%%%%%%%%%%%%%%%%%%%%%%%%%%%%%%%%%%%%%%%%%%%%%%%
\author[0000-0001-6143-8151]{Ricardo Carrera}
\affiliation{Astronomical Observatory of Padova, National Institute of Astrophysics, Vicolo Osservatorio 5 - 35122 - Padova}

% %%%%%%%%%%%%%%%%%%%%%%%%%%%%%%%%%%%%%%%%%%%%%%%%%%%%%%%%%%%%
\author{Cristina Chiappini}
\affiliation{Leibniz-Institut fur Astrophysik Potsdam (AIP), An der Sternwarte 16, 14482 Potsdam, Germany}

% %%%%%%%%%%%%%%%%%%%%%%%%%%%%%%%%%%%%%%%%%%%%%%%%%%%%%%%%%%%%
\author{Roger Cohen}
\affiliation{Space Telescope Science Institute, 3700 San Martin Drive, Baltimore, MD 21218, USA}

% %%%%%%%%%%%%%%%%%%%%%%%%%%%%%%%%%%%%%%%%%%%%%%%%%%%%%%%%%%%%
\author[0000-0002-1693-2721]{D. A. Garc{\'i}a-Hern{\'a}ndez}
\affiliation{Instituto de Astrof{\'i}sica de Canarias, V{\'i}a L\'actea S/N, 38205 La Laguna, Tenerife, Spain}
\affiliation{Universidad de La Laguna, Departamento de Astrof{\'i}sica, 30206 La Laguna, Tenerife, Spain}

% %%%%%%%%%%%%%%%%%%%%%%%%%%%%%%%%%%%%%%%%%%%%%%%%%%%%%%%%%%%%
\author{Doug Geisler}
\affiliation{Departamento de Astronom{\'{\i}}a, Universidad de Concepci{\'o}n, Casilla 160-C, Concepci{\'o}n, Chile}
\affiliation{Departamento de Astronom{\'{\i}}a, Universidad de La Serena, Avenida
Juan Cisternas 1200, La Serena, Chile}
\affiliation{Instituto de Investigaci{\'o}n Multidisciplinario en Ciencia y Tecnolog{\'{\i}}a,
Universidad de La Serena. Benavente 980, La Serena, Chile}

% %%%%%%%%%%%%%%%%%%%%%%%%%%%%%%%%%%%%%%%%%%%%%%%%%%%%%%%%%%%%
\author{Sten Hasselquist}
\altaffiliation{NSF Astronomy and Astrophysics Fellow}
\affiliation{Department of Physics \& Astronomy, University of Utah, 115 S. 1400 E., Salt Lake City, UT 84112, USA}

% %%%%%%%%%%%%%%%%%%%%%%%%%%%%%%%%%%%%%%%%%%%%%%%%%%%%%%%%%%%%
\author[0000-0002-4912-8609]{Henrik J\"onsson}
\affil{Materials Science and Applied Mathematics, Malm\"o University, SE-205 06 Malm\"o, Sweden}
\affil{Lund Observatory, Department of Astronomy and Theoretical Physics, Lund University, Box 43, SE-22100 Lund, Sweden}

% %%%%%%%%%%%%%%%%%%%%%%%%%%%%%%%%%%%%%%%%%%%%%%%%%%%%%%%%%%%%
\author{Richard R. Lane}
\affiliation{Instituto de Astrof\'isica, Pontificia Universidad Cat\'olica de Chile, Av. Vicuna Mackenna 4860, 782-0436 Macul, Santiago, Chile}
\affiliation{Instituto de Astronom\'ia y Ciencias Planetarias, Universidad de Atacama, Copayapu 485, Copiap\'o, Chile} 

% %%%%%%%%%%%%%%%%%%%%%%%%%%%%%%%%%%%%%%%%%%%%%%%%%%%%%%%%%%%%
\author{Steven R. Majewski}
\affil{Department of Astronomy, University of Virginia, Charlottesville, VA 22904-4325, USA}

% %%%%%%%%%%%%%%%%%%%%%%%%%%%%%%%%%%%%%%%%%%%%%%%%%%%%%%%%%%%%
\author[0000-0002-7064-099X]{Dante Minniti}
\affiliation{Departamento de Ciencias Fisicas, Facultad de Ciencias Exactas, Universidad Andres Bello, Av. \\Fernandez Concha 700, Las Condes, Santiago, Chile}
\affiliation{Millennium Institute of Astrophysics, Av. Vicuna Mackenna 4860, 782-0436, Santiago, Chile}
\affiliation{Vatican Observatory, V00120 Vatican City State, Italy}

% %%%%%%%%%%%%%%%%%%%%%%%%%%%%%%%%%%%%%%%%%%%%%%%%%%%%%%%%%%%%
\author{Christian Moni Bidin}
\affiliation{Instituto de Astronom\'ia, Universidad Cat\'olica del Norte, Av. Angamos 0610, Antofagasta, Chile}

% %%%%%%%%%%%%%%%%%%%%%%%%%%%%%%%%%%%%%%%%%%%%%%%%%%%%%%%%%%%%
\author[0000-0002-2835-2556]{Kaike Pan}
\affiliation{Apache Point Observatory and New Mexico State
    University, P.O. Box 59, Sunspot, NM, 88349-0059, USA} %(kpan, dmbiz@apo.nmsu.edu)}

% %%%%%%%%%%%%%%%%%%%%%%%%%%%%%%%%%%%%%%%%%%%%%%%%%%%%%%%%%%%%
\author[0000-0002-1379-4204]{Alexandre Roman-Lopes}
\affiliation{Department of Astronomy - Universidad de La Serena - Av. Juan Cisternas, 1200 North, La Serena, Chile}% - aroman@userena.cl}

% %%%%%%%%%%%%%%%%%%%%%%%%%%%%%%%%%%%%%%%%%%%%%%%%%%%%%%%%%%%%
\author[0000-0002-4989-0353]{Jennifer S. Sobeck}
\affiliation{Department of Astronomy, University of Washington, Seattle, WA, 98195, USA}

% %%%%%%%%%%%%%%%%%%%%%%%%%%%%%%%%%%%%%%%%%%%%%%%%%%%%%%%%%%%%
 \author[0000-0001-6761-9359]{Gail Zasowski}
 \affiliation{Department of Physics \& Astronomy, University of Utah, 115 S. 1400 E., Salt Lake City, UT 84112, USA}

% %%%%%%%%%%%%%%%%%%%%%%%%%%%%%%%%%%%%%%%%%%%%%%%%%%%%%%%%%%%%

%\author{APOGEE TEAM}
%\affiliation{SDSS APOGEE TEAM}

%%%%%%%%%%%%%%%%%%%%%%%%%%%%%%%%%%%%%%%%%%%%%%%%%%%%%%%%%%%%
\begin{abstract}

The Open Cluster Chemical Abundances and Mapping (OCCAM) survey aims  to constrain key Galactic dynamical and chemical evolution parameters by the construction of a large, comprehensive, uniform, infrared-based spectroscopic data set of hundreds of open clusters.
This fourth contribution from the OCCAM survey presents analysis  {using SDSS/APOGEE DR16} of a sample of 128 open clusters,  {71} of which we designate to be ``high quality'' based on the appearance of their color-magnitude diagram. We find the APOGEE DR16 derived [Fe/H] abundances to be in good agreement with previous high resolution  {spectroscopic} open cluster abundance studies.
Using the high quality sample, we measure Galactic abundance gradients in 16 elements, and find evolution of some of the [X/Fe] gradients as a function of age. We find an overall Galactic [Fe/H] vs \rgc gradient of $-0.068 \pm 0.001$ dex kpc$^{-1}$ over the range of $6 <$ \rgc $< 13.9$ kpc; however, we note that this result is sensitive to the distance catalog used,  {varying} as much as 15\%. We formally  {derive} the location a break in the [Fe/H] abundance gradient as a free parameter in the gradient fit for the first time. We also measure significant Galactic gradients in O, Mg, S, Ca, Mn, Cr, Cu, Na, Al, and K, some of which are measured for the first time. Our large sample allows us to explore four well-populated age bins to explore the time evolution of gradients for a large number of elements and comment on possible implications for Galactic chemical evolution and radial migration. 
\end{abstract}

%%=====================================================================================
%\clearpage
\section{ INTRODUCTION }

In this era of multi-fiber spectrographs, studies of tens of thousands of stars across the Galaxy are common. 
 {However, to} derive  {critical} parameters such as age and distance, the importance of reliable calibration samples cannot be understated. Open clusters serve as reliable age, distance, and chemical tracers distributed around the  {Galactic disk}.

Open clusters have been used to study Galactic chemical trends as far back as \citet{janes_79}, where the author showed open clusters to be a reliable tracer of a Galactic radial metallicity gradient. More recently, this trend has been consistently considered a 2-function gradient \citep[e.g.,][]{sestito2008, bragaglia2008, friel2010, carrera_2011, yong_2012, frinchaboy_13, reddy_16, magrini_2017}, with the break falling between $R_{GC} \approx10$ kpc and $R_{GC} \approx 16$ kpc. This gradient has become an important observable constraint for models of Galactic Chemical Evolution. Recent work has measured the inner gradient to be between $-0.05$ dex kpc$^{-1}$ \citep{reddy_16, casamiquela_2019}, and $-0.1$ dex kpc$^{-1}$ \citep{jacobson_16}. In addition, \citet{donor_18} (henceforth OCCAMII) showed that this gradient could change by as much as 40\% depending on which distance catalog was used. 

Since open clusters can range in age from a few Myr to more than 6 Gyr, they also provide a unique opportunity to study the evolution of  {G}alactic abundance gradient {s}. A number of authors have measured metallicity gradients for open clusters in various age bins \citep[e.g.,][]{carraro_98, friel_02, jacobson_2011, carrera_2011, cunha_16_grads}, and while all studies agree that the gradient is shallower for younger clusters, further comparison is difficult due to a somewhat heterogeneous choice of age bins; there does not seem to be a consensus as to the measured gradient for clusters of any given age range. 

Indeed, there are indications the picture is even more complicated. While open clusters have the advantage of precise age estimates, there are complexities that must be considered when using them to probe Galactic evolution. \citet{anders_2017} suggest open clusters in the inner galaxy are more likely to be broken up, leading to  {samples} significantly biased  {towards younger} clusters. 

Galactic trends in  {elements besides iron} have been reported \citep[e.g.,][]{yong_2005, friel2010, jacobson_2011}. Trend lines are commonly fit for $\alpha$-elements \citep[e.g.,][]{carrera_2011, yong_2012, reddy_16}, and in some cases for other elements {, such as [Ni/Fe], [Cr/Fe], and [V/Fe] \citep{casamiquela_2019} or [Na/Fe] and [Al/Fe] \citep{yong_2012}}. There is a growing consensus that there is a mild positive [$\alpha$/Fe]  {versus} \rgc trend in the inner galaxy, similar to some chemodynamical model predictions \citep[see][]{minchev_2}. OCCAMII showed the value of studying trends in other elements, finding strong evidence for a negative trend in [Mn/Fe] vs $R_{GC}$.

In this paper we will present the expanded OCCAM sample based on results from SDSS IV Apache Point Observatory Galactic Evolution Experiment 2 \citep[APOGEE 2;][]{apogee}  {Data Release 16 (DR16)} (J\"onsson et al., \textit{in prep}). We discuss this sample in comparison to the previously studied sample of open clusters that used SDSS IV DR14 results (OCCAMII), as well as other results from the literature. We then explore Galactic trends in [Fe/H], $\alpha$ elements, iron-peak elements, and all other elements reported by APOGEE as a function of  {G}alactocentric distances. We finally break the sample in age bins to explore changes in radial abundance trends over time.

%%%%%%%%%%%%%%%%%%%%%%%%%%%%%%%%%%%%%%%%%%%%%%%%%%%%%%%%%%%%%%%%%%%%%%%%%%%%%%%%%%%%%%%%%%%%%%%%%

\section{Data}

To minimize the impact of calibration differences and  {other} systematic effects,  {and ensure uniformity}, the OCCAM survey uses as much data from as few sources as possible; therefore, the majority of this analysis is based primarily on two large surveys, \gaia and SDSS/APOGEE.   

Our primary source of chemical abundance and radial velocity (RV) data is the Sloan Digital Sky Survey's (SDSS) sixteenth data release (DR16) \citep[Ahumada et al., {\it submitted}; J\"onsson et al., \textit{in prep};][]{sdss4} taken as part of  {the second, dual hemisphere phase of APOGEE (APOGEE 2)} \citep{apogee}. APOGEE is a high resolution,  {near} infrared spectroscopic survey currently operating in both hemispheres, at Apache Point Observatory \citep[APO; New Mexico,][]{sloan_telescope} and Las Campanas Observatory \citep[LCO; Chile,][]{du_pont}.  The APOGEE/DR16 dataset includes about 430,000 stars, collected between August 2011 and August 2018 using the two 300-fiber APOGEE spectrographs \citep{wilson_2019} and, for the first time, the APOGEE survey has near-complete coverage in Galactic longitude, due to the first release of data from LCO.  The APOGEE data reduction pipeline \citep[][J\"onsson et al., \textit{in prep}]{nidever_2015, holtzman_2015, holtzman_2018} provides stellar  {atmospheric} parameters and radial velocity measurements,  {while} elemental abundances are provided from the ASPCAP pipeline \citep[][J\"onsson et al., \textit{in prep}]{aspcap, meszaros_2012, zamora_15, holtzman_2018}. Copper, cerium \citep{cunha_2017},  neodymium \citep{Hasselquist_2016}, and ytterbium abundances are reported from ASPCAP for the first time in DR16, although neodymium and ytterbium lines are so weak or blended that these  {ASPCAP} abundances are considered unreliable. 
Concerning cerium, the APOGEE region contains several Ce II lines \citep{cunha_2017}, however, the current DR16 results are only based on one Ce II line; future data releases will use the full sample of cerium lines.  {Therefore we will postpone any discussion of cerium until future data releases.}

 {In the APOGEE DR16 allStar-file several types of abundances are reported for every star and element: firstly the abundance reported by the analysis pipeline is supplied in the FELEM-array. Secondly, these abundances have been
calibrated with a zero-point shift to ensure solar metallicity stars in the solar neighborhood have [X/M]=0; in practice these shifts are small, $< 0.05$ dex, except for Al, K, V, and Mn.
Finally, these calibrated abundances have been culled for particular uncertain values by the ASPCAP-team (e.g., for [Y/Fe] or [Nd/Fe]). These final, ``cleaned'' and calibrated abundances are supplied in the ``named tags''; FE\_H, MG\_FE, CE\_FE, etc. More information, including what zero-point shifts have been applied, is provided in J\"onsson et al., \textit{in prep}. In this paper we use the abundances of the ``named tags'' as is recommended in J\"onsson et al., \textit{in prep}.}

Targeting for APOGEE relied on input from two all-sky surveys: 2MASS \citep{2MASS_DR_allsky} and WISE \citep{WISE_mission_description}. More details  {specifically} about open cluster targeting are provided in OCCAMII, and details about APOGEE targeting  {generally} can be found in \citet{zasowski13, zasowski17}.
%, Beaton et al., \textit{in prep}, and Santana et al., \textit{in prep}.

Our secondary source of data is \gaia DR2 \citep{gaia_mission, gaia_dr2, gaia_astrometry}; we use photometric and astrometric data for 1,365,376 \gaia stars, radial velocity  {measurements} for 16,084 stars, and parallax values for 886 stars in common with APOGEE. 

We use  {cluster} coordinates and radii from \citet{dias_catalog}. 
For this study, we use the uniform distance determination from \citet[generally referred to as the Milky Way Star Cluster, MWSC, catalog]{mwsc_catalog} when measuring galactic trends; however, we briefly compare to other uniform distance catalogs \citep[e.g.,][]{cg_18, bailer_jones_cat} in \S \ref{sec:distProb}. 

\begin{deluxetable*}{lrrrrrrrrrrr}[b!]
\tabletypesize{\scriptsize}
\tablecaption{OCCAM DR16 Sample - Basic Parameters \label{tab:full_sample}}
	\tablehead{
    \colhead{Cluster} &
    \colhead{Qual} &
    \colhead{l} &
    \colhead{b} &
    \colhead{R\tablenotemark{a}} &
     \colhead{Age\tablenotemark{b}} &
    \colhead{R$_{GC}$\tablenotemark{b}} & 
    \colhead{$\mu_{\alpha}$\tablenotemark{c}} &
    \colhead{$\mu_{\delta}$\tablenotemark{c}} &
    \colhead{RV} & 
    % \colhead{$\sigma_{RV}$} & 
    \colhead{[Fe/H]} & 
    % \colhead{$\sigma_{[Fe/H]}$} & 
    \colhead{Num}\\[-2ex] %} & 
    %\colhead{APOGEE}\\[-2ex]
    \colhead{name} &
    \colhead{flag} &
    \colhead{deg} &
    \colhead{deg} &
    \colhead{(\arcmin)} &
     \colhead{Gyr} &
    \colhead{(kpc)} &
    \colhead{(mas yr$^{-1}$)} &
    \colhead{(mas yr$^{-1}$)} &
    % \colhead{(km s$^{-1}$)} & 
    \colhead{(km s$^{-1}$)} & 
    \colhead{(dex)} &
    % \colhead{(dex)} &
    \colhead{stars} %& 
    %\colhead{stars\tablenotemark{c}}
    }
	\startdata
\multicolumn{11}{c}{High Quality Clusters}\\\hline
Ruprecht 147        &  1 & 21.0089 &  $-$12.7301 & 30.0 & 2.14 & 7.72 & $-$0.87 $\pm$ 0.10 & $-$26.72 $\pm$ 0.10 & $+$42.4 $\pm$ 1.5 & $+$0.12 $\pm$ 0.03 &  27\\
NGC 6705            &  1 & 27.2873 &  $-$2.7594 & 9.0 & 0.32 & 5.94 & $-$1.56 $\pm$ 0.08 & $-$4.17 $\pm$ 0.07 & $+$35.4 $\pm$ 1.0 & $+$0.12 $\pm$ 0.04 &  12\\
Berkeley 43         &  1 & 45.6843 &  $-$0.1391 & 6.3 & 0.61 & 5.73 & $-$0.92 $\pm$ 0.08 & $-$3.27 $\pm$ 0.07 & $+$30.0 $\pm$ 0.1 & $+$0.03 $\pm$ 0.01 &   1\\
Berkeley 44         &  1 & 53.2093 &  $+$3.3443 & 6.3 & 1.41 & 6.50 & $-$0.17 $\pm$ 0.05 & $-$3.17 $\pm$ 0.05 & $+$23.0 $\pm$ 0.1 & $-$0.00 $\pm$ 0.01 &   1\\
NGC 6791            &  2 & 69.9658 &  $+$10.9080 & 6.3 & 4.42 & 7.71 & $-$0.44 $\pm$ 0.03 & $-$2.25 $\pm$ 0.03 & $-$46.9 $\pm$ 1.3 & $+$0.35 $\pm$ 0.04 &  36\\
NGC 6819            &  2 & 73.9834 &  $+$8.4882 & 6.9 & 1.62 & 7.70 & $-$2.96 $\pm$ 0.03 & $-$3.87 $\pm$ 0.03 & $+$2.7 $\pm$ 1.7 & $+$0.05 $\pm$ 0.03 &  37\\
NGC 6811            &  2 & 79.2233 &  $+$12.0047 & 7.2 & 0.64 & 7.87 & $-$3.44 $\pm$ 0.06 & $-$8.73 $\pm$ 0.04 & $+$8.0 $\pm$ 0.3 & $-$0.05 $\pm$ 0.02 &   4\\
NGC 6866            &  1 & 79.5648 &  $+$6.8354 & 5.1 & 0.44 & 7.87 & $-$1.18 $\pm$ 0.04 & $-$5.91 $\pm$ 0.08 & $+$14.2 $\pm$ 0.4 & $+$0.01 $\pm$ 0.01 &   2\\
IC 1369             &  1 & 89.6019 &  $-$0.4154 & 5.1 & 0.35 & 8.70 & $-$4.68 $\pm$ 0.05 & $-$5.55 $\pm$ 0.04 & $-$48.5 $\pm$ 0.1 & $-$0.08 $\pm$ 0.03 &   3\\
NGC 7062            &  1 & 89.9667 &  $-$2.7397 & 3.6 & 0.69 & 8.34 & $-$1.84 $\pm$ 0.04 & $-$4.08 $\pm$ 0.04 & $-$22.0 $\pm$ 0.1 & $+$0.01 $\pm$ 0.01 &   1\\
\multicolumn{11}{c}{......}
\enddata
\tablenotetext{a}{Radius from \citet{dias_catalog}}\vskip-0.07in
\tablenotetext{b}{Calculated using or taken from MWSC Catalog.}\vskip-0.07in
\tablenotetext{c}{$\mu_{\alpha}$ and $\mu_{\delta}$ and their $1 \sigma $ uncertainties are those of the 2D Gaussian fit, as in OCCAMII.}\vskip-0.07in
\tablenotetext{}{(This table is available in its entirety in machine-readable form.)}
\end{deluxetable*}

\begin{deluxetable*}{lrrrrrrrr}
\tabletypesize{\tiny}
\tablecaption{OCCAM DR16 Sample - Detailed Chemistry \label{tab:full_sample2}}
	\tablehead{
    \colhead{Cluster} & 
    \colhead{[Fe/H]} &
    \colhead{[O/Fe]} & 
    \colhead{[Na/Fe]} & 
    \colhead{[Mg/Fe]} & 
    \colhead{[Al/Fe]} & 
    \colhead{[Si/Fe]} & 
    \colhead{[S/Fe]} &
    \colhead{[K/Fe]} \\[-5ex]
     \colhead{name} &
     \colhead{(dex)} &
     \colhead{(dex)} &
     \colhead{(dex)} &
     \colhead{(dex)} &
     \colhead{(dex)} &
     \colhead{(dex)} &
     \colhead{(dex)} &
     \colhead{(dex)} \\[-2ex]
    \colhead{} &
    \colhead{[Ca/Fe]} &
    \colhead{[Ti/Fe]} &
    \colhead{[V/Fe]} &
    \colhead{[Cr/Fe]} & 
    \colhead{[Mn/Fe]} & 
    \colhead{[Co/Fe]} & 
    \colhead{[Ni/Fe]} & 
    \colhead{[Cu/Fe]}\\[-5ex] 
     \colhead{} &
     \colhead{(dex)} &
     \colhead{(dex)} &
     \colhead{(dex)} &
     \colhead{(dex)} &
     \colhead{(dex)} &
     \colhead{(dex)} &
     \colhead{(dex)} &
     \colhead{(dex)}
    }
\startdata
\multicolumn{9}{c}{High Quality Clusters} \\ \hline
Ruprecht 147         & $0.12 \pm 0.03$ &$-0.05 \pm 0.03$ &$0.11 \pm 0.03$ &$-0.01 \pm 0.02$ &$0.02 \pm 0.04$ &$-0.00 \pm 0.05$ &$0.02 \pm 0.06$ &$0.04 \pm 0.08$ \\& $-0.01 \pm 0.04$ &$-0.07 \pm 0.09$ &$0.01 \pm 0.07$ &$0.02 \pm 0.09$ &$0.04 \pm 0.03$ &$0.14 \pm 0.20$ &$0.01 \pm 0.02$ &$-0.09 \pm 0.20$ \\[1ex]
NGC 6705             & $0.12 \pm 0.04$ &$-0.05 \pm 0.02$ &$0.23 \pm 0.04$ &$-0.07 \pm 0.02$ &$-0.13 \pm 0.03$ &$0.01 \pm 0.01$ &$0.07 \pm 0.02$ &$-0.16 \pm 0.06$ \\& $-0.03 \pm 0.02$ &$-0.00 \pm 0.02$ &$-0.01 \pm 0.04$ &$-0.03 \pm 0.04$ &$0.11 \pm 0.02$ &$0.04 \pm 0.03$ &$0.03 \pm 0.01$ &$0.17 \pm 0.07$\\[1ex]
Berkeley 43          & $0.03 \pm 0.01$ &$-0.05 \pm 0.01$ &$0.15 \pm 0.03$ &$-0.08 \pm 0.01$ &$-0.22 \pm 0.02$ &$0.03 \pm 0.01$ &$0.13 \pm 0.02$ &$-0.18 \pm 0.03$ \\& $-0.05 \pm 0.01$ &$-0.01 \pm 0.01$ &$-0.02 \pm 0.04$ &$-0.08 \pm 0.03$ &$0.12 \pm 0.01$ &$0.01 \pm 0.03$ &$0.02 \pm 0.01$ &$-0.26 \pm 0.03$\\[1ex]
Berkeley 44          & $-0.00 \pm 0.01$ &$0.04 \pm 0.01$ &$-0.16 \pm 0.03$ &$-0.02 \pm 0.01$ &$-0.30 \pm 0.02$ &$0.01 \pm 0.01$ &$0.02 \pm 0.03$ &$-0.15 \pm 0.03$ \\& $-0.14 \pm 0.01$ &$-0.13 \pm 0.01$ &$-0.29 \pm 0.04$ &$-0.17 \pm 0.03$ &$0.06 \pm 0.01$ &$0.05 \pm 0.03$ &$-0.03 \pm 0.01$ &$0.04 \pm 0.03$\\[1ex]
NGC 6791             & $0.35 \pm 0.04$ &$0.04 \pm 0.03$ &$0.08 \pm 0.06$ &$0.11 \pm 0.03$ &$0.01 \pm 0.07$ &$0.01 \pm 0.03$ &$-0.02 \pm 0.05$ &$0.03 \pm 0.10$ \\& $-0.02 \pm 0.03$ &$0.09 \pm 0.05$ &$-0.06 \pm 0.30$ &$-0.02 \pm 0.09$ &$0.01 \pm 0.13$ &$0.11 \pm 0.08$ &$0.01 \pm 0.04$ &$0.14 \pm 0.07$\\[1ex]
NGC 6819             & $0.05 \pm 0.03$ &$-0.01 \pm 0.03$ &$0.07 \pm 0.09$ &$-0.01 \pm 0.01$ &$-0.04 \pm 0.03$ &$0.00 \pm 0.03$ &$0.00 \pm 0.03$ &$-0.04 \pm 0.07$ \\& $0.01 \pm 0.02$ &$0.01 \pm 0.03$ &$0.04 \pm 0.13$ &$0.01 \pm 0.03$ &$0.05 \pm 0.03$ &$0.02 \pm 0.06$ &$0.02 \pm 0.02$ &$0.04 \pm 0.06$\\[1ex]
NGC 6811             & $-0.05 \pm 0.02$ &$-0.04 \pm 0.02$ &$0.06 \pm 0.07$ &$-0.04 \pm 0.01$ &$-0.07 \pm 0.03$ &$-0.02 \pm 0.01$ &$0.05 \pm 0.04$ &$-0.06 \pm 0.05$ \\& $0.02 \pm 0.01$ &$0.00 \pm 0.02$ &\multicolumn{1}{c}{\nodata} &$0.05 \pm 0.03$ &$0.01 \pm 0.02$ &$-0.16 \pm 0.12$ &$-0.03 \pm 0.01$ &$-0.06 \pm 0.10$\\[1ex]
NGC 6866             & $0.01 \pm 0.01$ &\multicolumn{1}{c}{\nodata} &$-0.00 \pm 0.04$ &$-0.05 \pm 0.01$ &$-0.04 \pm 0.02$ &$-0.04 \pm 0.01$ &$0.04 \pm 0.03$ &$-0.06 \pm 0.03$ \\& $0.01 \pm 0.02$ &$0.01 \pm 0.02$ & \multicolumn{1}{c}{\nodata} &$0.03 \pm 0.05$ &$0.02 \pm 0.01$ &$-0.14 \pm 0.08$ &$-0.03 \pm 0.01$ &$0.02 \pm 0.03$\\[1ex]
IC 1369              & $-0.08 \pm 0.03$ &$-0.08 \pm 0.02$ &$0.08 \pm 0.10$ &$-0.04 \pm 0.02$ &$-0.11 \pm 0.02$ &$-0.01 \pm 0.01$ &$0.09 \pm 0.07$ &$0.02 \pm 0.03$ \\& $0.01 \pm 0.04$ &$-0.08 \pm 0.02$ & \multicolumn{1}{c}{\nodata} &$0.01 \pm 0.04$ &$0.04 \pm 0.03$ &$-0.04 \pm 0.04$ &$-0.06 \pm 0.01$ &$0.09 \pm 0.04$\\[1ex]
NGC 7062             & $0.01 \pm 0.01$ & \multicolumn{1}{c}{\nodata}  &$0.17 \pm 0.04$ &$-0.07 \pm 0.01$ &$-0.05 \pm 0.02$ &$-0.01 \pm 0.01$ &$0.00 \pm 0.03$ &$-0.05 \pm 0.03$ \\& $-0.00 \pm 0.01$ &$-0.02 \pm 0.02$ & \multicolumn{1}{c}{\nodata}  &$-0.08 \pm 0.03$ &$-0.00 \pm 0.01$ &$0.01 \pm 0.04$ &$-0.02 \pm 0.01$ &$-0.04 \pm 0.03$\\[1ex]
\multicolumn{9}{c}{......}\\
\enddata
\tablenotetext{}{(This table is available in its entirety in machine-readable form.)}
\end{deluxetable*}

%%%%%%%%%%%%%%%%%%%%%%%%%%%%%%%%%%%%%%%%%%%%%%%%%%%%%%%%%%%%%%%%%%%%%%%%%%%%%%%%%%%%%%%%%%%%%%%%%
\section{Methods}

\subsection{Membership Analysis}

The selection of cluster member stars utilizes the stellar radial velocities, proper motions (PM), spatial location, and derived metallicities as membership discriminators.
For this study, we use the membership procedure, fully described in OCCAMII with some minor improvements. 
The method of OCCAMII first performs a PM analysis using \gaia DR2 to isolate likely cluster members. If multiple APOGEE stars are selected  {for the same cluster that} have very different RVs, there is an inherent ambiguity and a ``correct''  {systemic cluster velocity} cannot be chosen. We now leverage the RV measurements from \gaia, when available, for stars identified as likely PM members to significantly increase the number of RV measurements in a cluster and  {more reliably determine the cluster system velocity}.

To be included as a cluster member, a star must fall within $3\sigma$ of the cluster mean as established by the kernel convolution (described in OCCAMII) in \textit{all three} spaces considered (RV, [Fe/H], and PM).

\subsection{Visual Quality Check}

\begin{figure*}
  \epsscale{1.15}
 \plotone{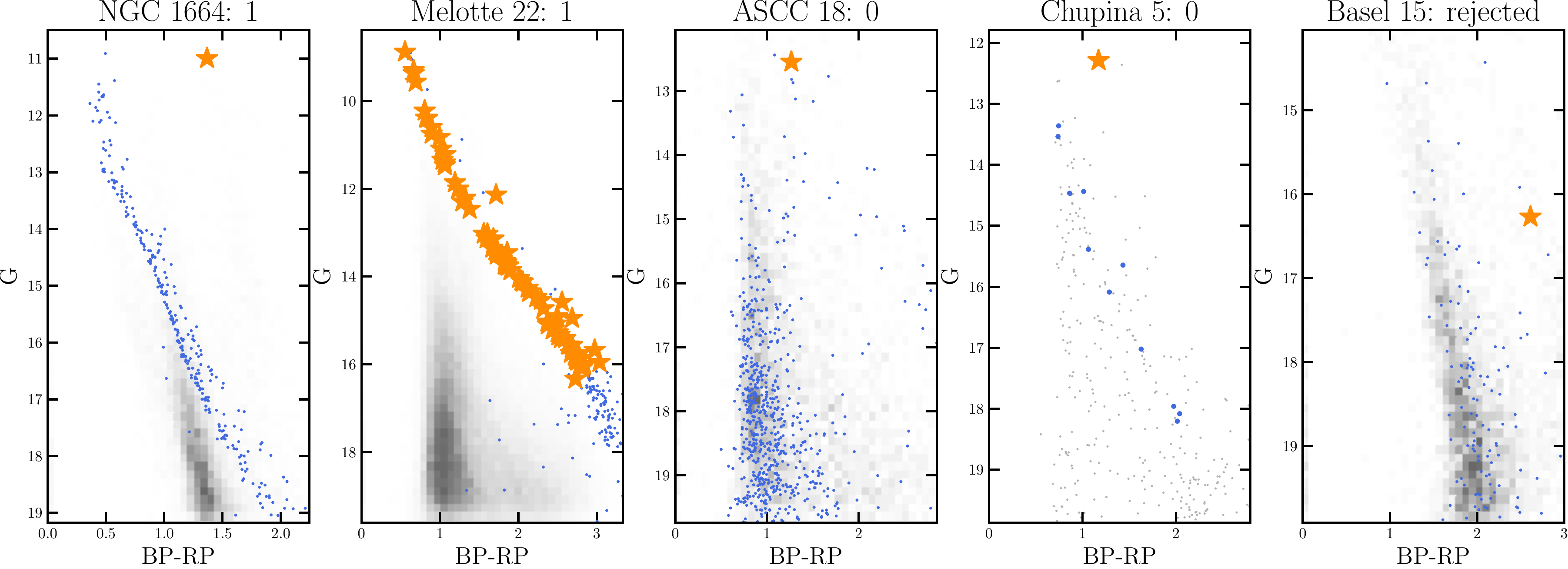}
 	\caption{ \small Five example color-magnitude diagrams of open clusters analyzed in the study, with cluster name  {and quality designation} from Table \ref{tab:full_sample} . \gaia stars  {within twice the cluster radius are shown; stars} identified as PM members  {and inside the cluster radius} are  {blue}.  {Non-member stars are shown as a Hess diagram in grey except for Chupina 5 where actual stars are shown}. The OCCAM pipeline-identified APOGEE members are shown as orange stars.}
 	\label{fig:qual}
 \end{figure*}

A visual inspection of each  {cluster's} PM-cleaned color-magnitude diagram (CMD) was performed by multiple of the authors. Figure \ref{fig:qual} shows five example CMDs. 
The visual assessment is meant to evaluate whether stars that pass the combined RV, proper motion and metallicity criteria also lie in a sensible position in the observed cluster CMD, considering their spectroscopically determined log(g).  This is an easy case when, for example, one or more APOGEE OCCAM candidates with high log(g) ($log(g) \ge 3.7$) are found to lie along an easily discernible photometric main sequence in the CMD (e.g., Melotte 22), thus providing a joint affirmation that the star is likely a main sequence member of the cluster. These clusters are flagged as ``1'' or ``high quality''.  However, most of the OCCAM stars from APOGEE turn out to be evolved stars -- subgiants, giants and red clump stars, with $log(g) <3.7$.  In this case, the star is still considered a member if the star lies along the subgiant/giant branch of the cluster, which, however, must generally be projected from the location of the main sequence and its turn-off, given that the subgiant/giant sequences in most clusters are typically very poorly populated (e.g., NGC 1664). These clusters are also flagged as ``1'' or ``high quality''.    The latter process becomes more challenging when the main sequence is also poorly populated (e.g., Chupina 5), or when the field star contamination becomes so dominant as to obscure the cluster main sequence (e.g., ASCC 18). These clusters are flagged as ``0'' or ``potentially unreliable''. Clusters where the APOGEE OCCAM candidate is not a part of any discernable sequence or where there is no discernable sequence (e.g., Basel 15) are rejected.
These quality flags are included in the full version of Tables \ref{tab:full_sample} {and} \ref{tab:full_sample2} (available online), and in the value added catalog, described below.

\subsection{Data Access - SDSS Value Added Catalog}

The data this analysis uses are also available as a Value Added Catalog (VAC) that was released along with SDSS-IV DR16. The VAC consists of two tables. The first is a combination of Table \ref{tab:full_sample} and Table \ref{tab:full_sample2}, showing bulk cluster parameters derived here including PM, and RV, but also including abundances for all\footnote{Elements such as Rb and Y  {that} do not have calibrated values reported in DR16 are not included.} elements reported in DR16. 
We note that cluster ages are not included in the VAC as only ages from the MWSC catalog are used in this work.

Five measurements of $R_{GC}$ are also included. We calculate \rgc using catalog distances from \citet{dias_catalog}\footnote{We acknowledge an error in our pipeline that populated \rgc for some clusters where no distance is reported by \citet{dias_catalog}. "R\_GC\_DIAS" values for the clusters ASCC 16, Chupina 3, 4, \& 5, Collinder 95, FSR 0687, L 1241s, NGC 358, and Platais 4 should be disregarded}, \citet[MWSC]{mwsc_catalog}, and \citet{cg_18}. We also calculate \rgc based on median parallax from member stars and median distance for member stars from \citet{bailer_jones_cat}, as in OCCAMII. In \S \ref{sec:distProb} we discuss differences in these distance measurements.

The second table in the VAC shows all of the APOGEE stars considered in this analysis (all the stars that fall within $2 \times Radius_{Dias}$ of the cluster center). For each star, we reproduce relevant parameters (RV, [Fe/H], and proper motion) and provide our membership probability estimate based on each parameter. For convenience, we also provide the membership determination from \citet{cg_18} (when provided).  All columns available in the VAC are presented in Table 3. The catalog is available  {from sdss.org} \href{https://www.sdss.org/dr16/data_access/value-added-catalogs/?vac_id=open-cluster-chemical-abundances-and-mapping-catalog}{here}\footnote{The full url is \url{https://www.sdss.org/dr16/data_access/value-added-catalogs/?vac_id=open-cluster-chemical-abundances-and-mapping-catalog}}.

Both tables are also available for exploration using Filtergraph \citep{filtergraph}  {at} \url{https://filtergraph.com/sdss_apogee_occam/}.

\begin{deluxetable}{ll}
\tablecaption{A summary of the individual star data included in the DR16 OCCAM VAC}
\tablehead{
    \colhead{Label} & 
    \colhead{Description}
    }
\startdata
CLUSTER & The associated open cluster \\
2MASS ID &   {star} ID from 2MASS survey\\
LOCATION\_ID\tablenotemark{a} & from APOGEE DR16 \\
GLAT &  Galactic latitude\\
GLON &  Galactic longitude\\
FE\_H\tablenotemark{a}  & [Fe/H]\\
FE\_H\_ERR\tablenotemark{a}  & uncertainty in FE\_H \\
VHELIO\_AVG\tablenotemark{a}  &  heliocentric radial velocity\\
VSCATTER\tablenotemark{a}  & scatter in APOGEE RV measurements \\
PMRA\tablenotemark{b} &  proper motion in right ascension\\
PMDEC\tablenotemark{b}  & proper motion in declination \\
PMRA\_ERR\tablenotemark{b}  & uncertainty in PMRA \\
PMDEC\_ERR\tablenotemark{b}  & uncertainty in PMDEC \\
RV\_PROB &  membership probability based on RV (This study)\\
FEH\_PROB &  membership probability based on FE\_H (This study)\\
PM\_PROB &  membership probability based on PM (This study)\\
CG\_PROB & membership probability from \citet{cg_18} \\
\enddata
\tablenotetext{a}{Taken directly from APOGEE DR16}\vskip-0.07in
\tablenotetext{b}{From \gaia DR2}
\end{deluxetable}\vskip-0.07in

%%%%%%%%%%%%%%%%%%%%%%%%%%%%%%%%%%%%%%%%%%%%%%%%%%%%%%%%%%%%%%%%%%%%%%%%%%%%%%%%%%%%%%%%%%%%%%%%%
%%%%%%%%%%%%%%%%%%%%%%%%%%%%%%%%%%%%%%%%%%%%%%%%%%%%%%%%%%%%%%%%%%%%%%%%%%%%%%%%%%%%%%%%%%%%%%%%%
%%%%%%%%%%%%%%%%%%%%%%%%%%%%%%%%%%%%%%%%%%%%%%%%%%%%%%%%%%%%%%%%%%%%%%%%%%%%%%%%%%%%%%%%%%%%%%%%%

\section{The OCCAM DR16 Sample}

Our final sample in this study consists of 128 open clusters with 914 member stars, out of 10,191 stars near cluster fields considered in the analysis. Of those 128 clusters, 83 clusters were designated as ``high quality'' based on a visual CMD inspection. 
For the Galactic abundance analysis in this study, we will only use those clusters flagged as high quality,  {as} presented in Table \ref{tab:full_sample}.
The other clusters with questionable quality, e.g., those that did not pass visual checks (\S3.2), are also presented in Table \ref{tab:full_sample}.

\begin{figure}
  \epsscale{1.2}
 \plotone{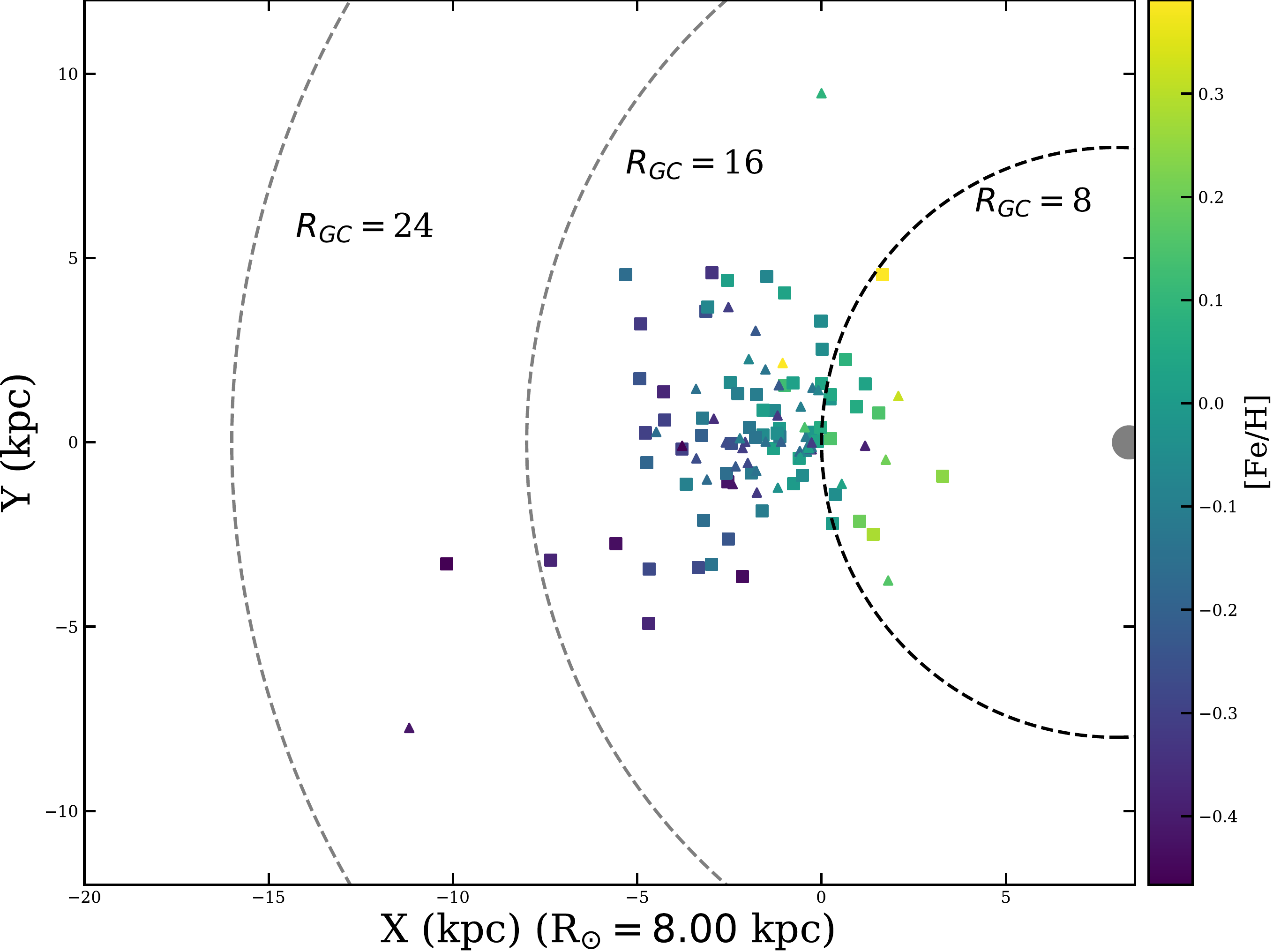}
 	\caption{ \small The full OCCAM DR16 sample plotted in the Galactic plane. Square points are ``high quality'' clusters, triangles are the lower quality clusters. The colorbar shows [Fe/H]. The concentric circles show $R_{GC}$ = 8, 16, \& 24 kpc }
 	\label{fig:XYfull}
 \end{figure}
%\clearpage

 {The Galactic spatial distribution of the OCCAM DR16 sample is shown in Figure \ref{fig:XYfull}.}
The majority of the OCCAM DR16 open clusters fall between $6 \le$ \rgc $\le 14$ kpc, with good \rgc coverage in that range. Two high quality clusters fall outside of this range: Berkeley 20 at \rgc $\approx 15.5$ kpc and Berkeley 29 at \rgc $\approx 18.5$ kpc\footnote{We note \citet{dias_catalog} find Be 29 to be significantly further away at \rgc $\approx 22.5$ kpc, but for consistency we are using distances from the MWSC catalog for all clusters.}.
Using age estimates from the MWSC catalog, our sample spans a range in age from $\sim 5$ Myr to $\sim 6$ Gyr\footnote{We note some studies of NGC 6791 \citep[e.g.][]{Brogaard_2012} find it to be significantly older, however in the interest of a uniform analysis we rely only on ages from the MWSC catalog}, with nearly half under 1 Gyr.

\subsection{Modifications to the High Quality Sample}
Beyond  {those clusters excluded from analysis based on our visual inspection of their PM-cleaned CMDs}, we have further excluded  {12 clusters (ASCC 16, ASCC 19, ASCC 21, Briceno 1, Chupina 1, Chupina 3, Collinder 69, Collinder 70, IC 348, NGC 1980, NGC 1981, NGC 2264)} because they are reported to be very young  {($<50$ Myr)} \citep{mwsc_catalog}  {and previous studies of young stars in APOGEE suggest the pipeline results may be unreliable \citep[e.g.,][]{kounkel_2018}.}
 {Thus the final sample used for analysis consists of 71 clusters.}

There are additional affects within clusters that may result in unreliable abundance determinations.  \citet{souto_2018, souto_19} showed that abundances in dwarf and giant stars in the old cluster NGC 2682 differed significantly due to atomic diffusion. For this reason, the dwarf stars in NGC 2682 are excluded from our abundance analysis. NGC 752 is also relatively old and may suffer from diffusion effects, we therefore exclude the dwarfs in this cluster from abundance analysis as well.  As a result, for both NGC 752 \& NGC 2682 we only use the giant stars to determine the cluster abundances.

%\clearpage
\subsection{Comparison to previous work}

 \begin{figure}
  \epsscale{1.2}
 \plotone{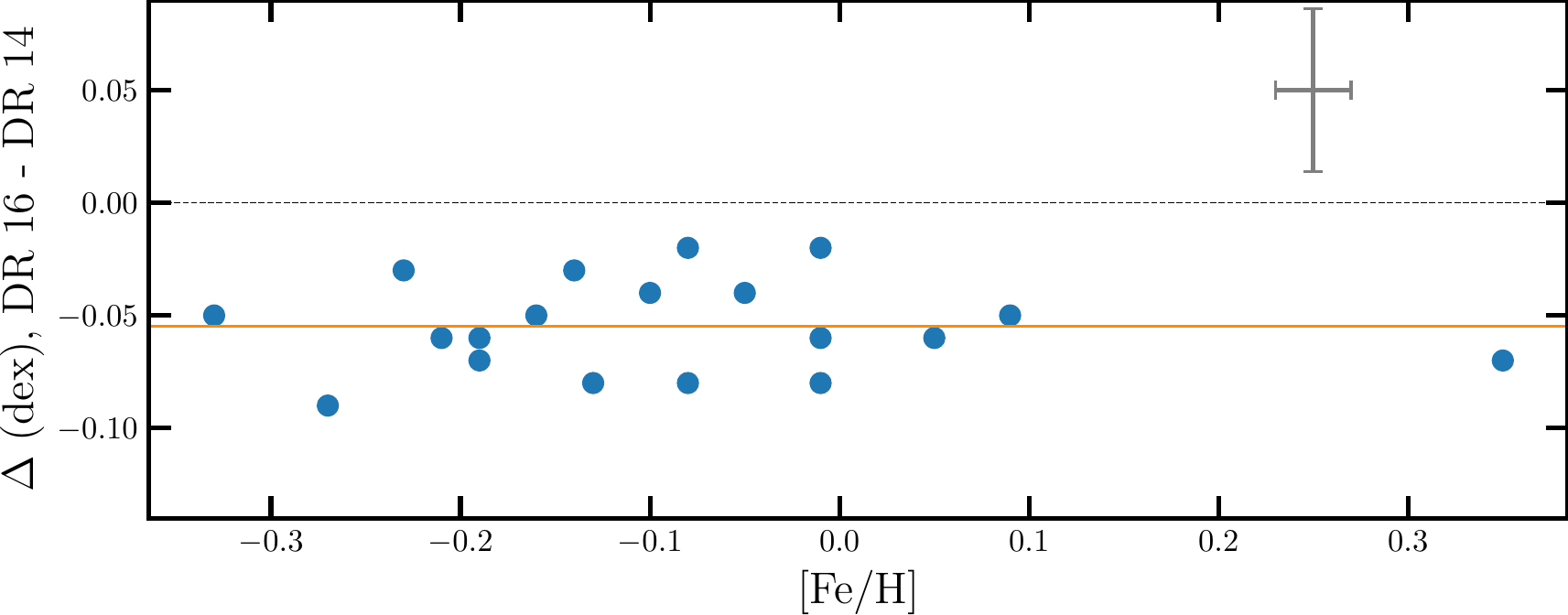}
 	\caption{ \small The difference in reported [Fe/H] from DR14 to DR16 for the 19 clusters from OCCAMII. A characteristic error-bar is shown.}
 	\label{fig:dr14v16_fe}
 \end{figure}
 
\subsubsection{OCCAM PAPER II}\label{sec:dr14comp}

 {For the 19 open clusters studied in OCCAMII, we plot $\Delta$ [Fe/H] vs DR16 [Fe/H] in Figure \ref{fig:dr14v16_fe}. } Figure \ref{fig:dr14v16_fe} shows that the mean [Fe/H] for OCCAM clusters changed between APOGEE DR14 and DR16; this is mostly due to changes in the gf-values of the Fe I lines in the DR16 line list (Smith et al. \textit{in prep}). There is a clear offset for all clusters, with a mean difference of 0.05 dex. In OCCAMII it was shown that APOGEE DR14 [Fe/H] values for six well studied open clusters were on average approximately 0.05 dex more metal-rich than the results in the literature. If we repeat the same literature comparison using our DR16 values we find a mean offset of [Fe/H] = 0.004. All of these offsets are within their measured $1 \sigma$ dispersions.

\begin{figure}
    \epsscale{1.2}
    \plotone{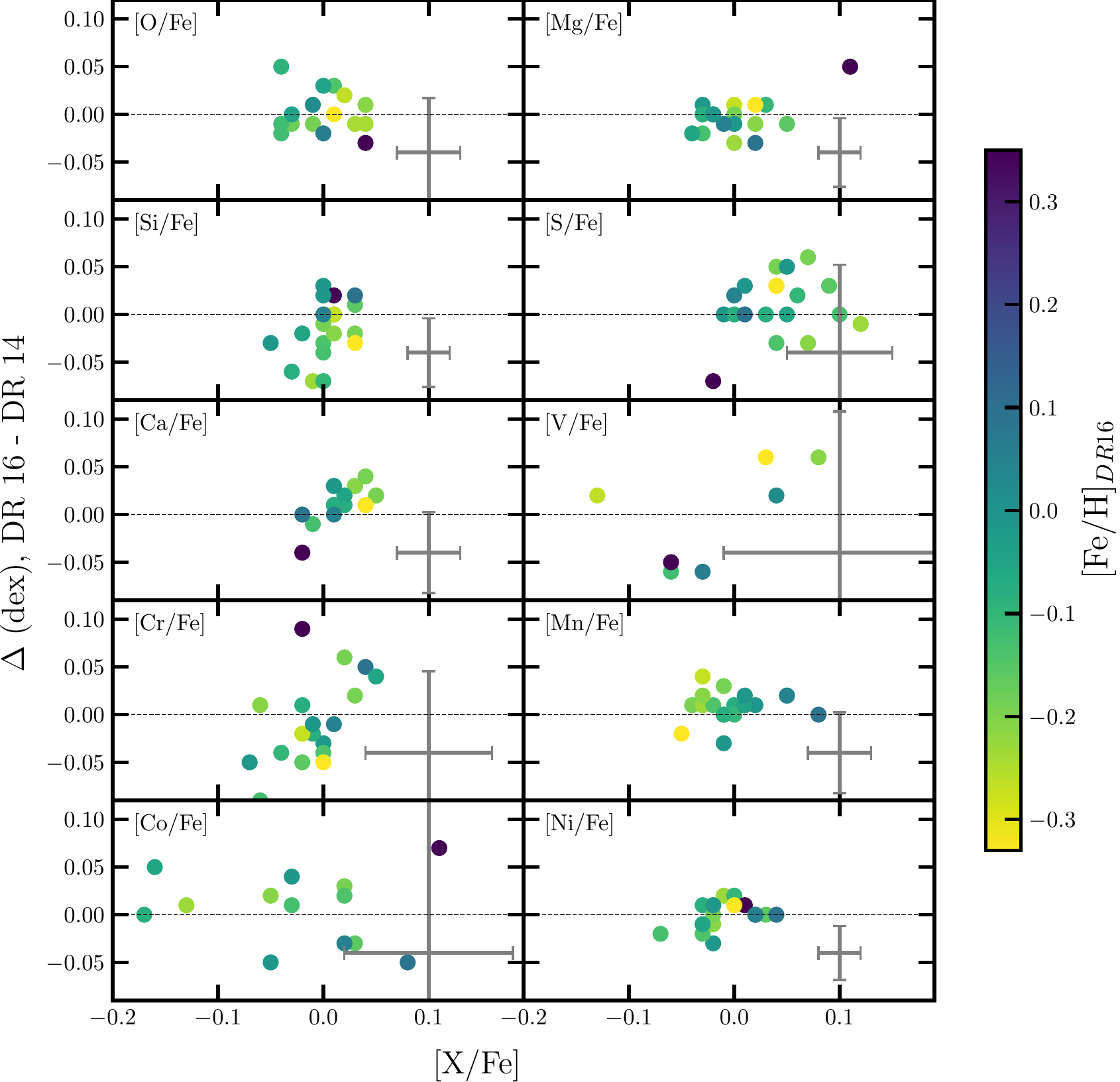}
 	%\vskip-0.35in
 	\caption{ \small Similar to Figure \ref{fig:dr14v16_fe} but for other elements. Characteristic error bars are shown. Datapoints are colored by  {their [Fe/H] as reported in APOGEE DR16}}
 	\label{fig:dr14v16}
 \end{figure}

Figure \ref{fig:dr14v16} shows a similar plot for other elements. Beyond the quoted uncertainties in each case, there are no obvious systematic trends for any of these elements.

%\clearpage
\subsubsection{Open Clusters Observed by the LAMOST Survey}

\begin{figure}[t]
 	\begin{center}
 		%\vskip-0.1in
 		%\includegraphics[width=\linewidth]
        \epsscale{1.2}
     \plotone{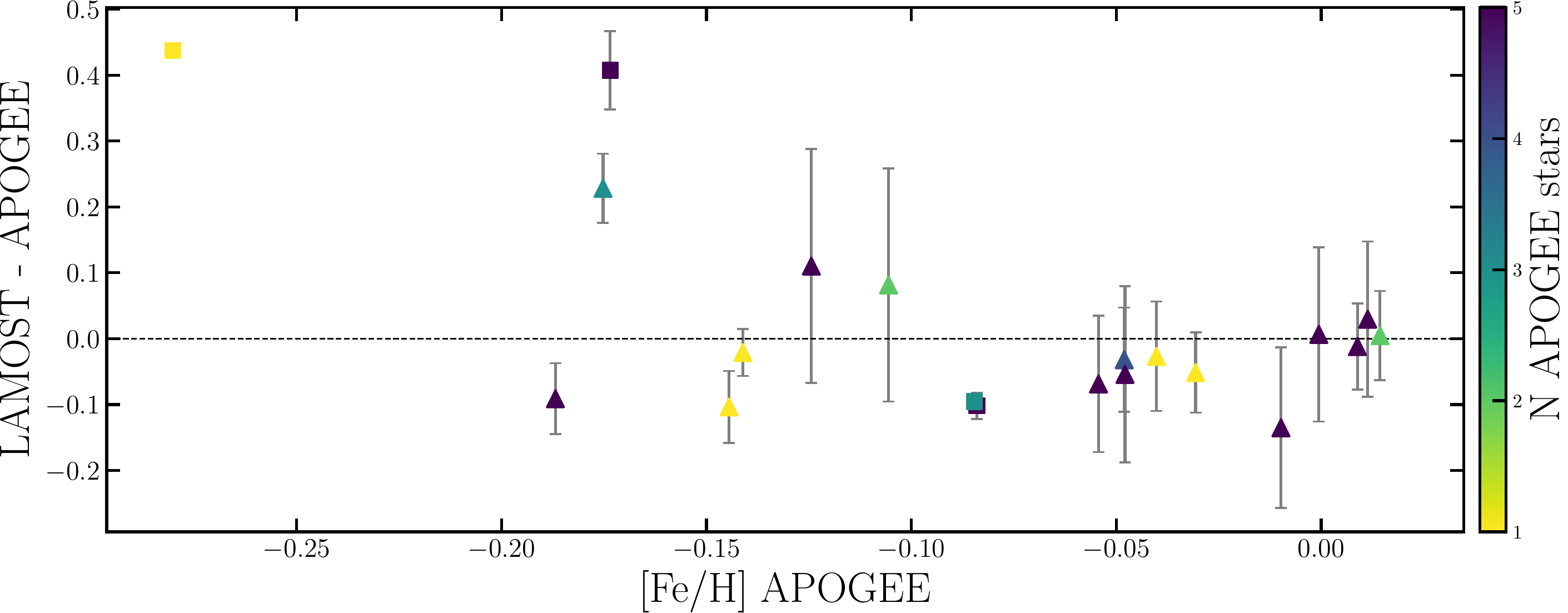}
 	\end{center}
 	%\vskip-0.35in
 	\caption{ \small The difference between the metallicities in the LAMOST \citep[from ][]{zhang_19} and APOGEE surveys for open clusters in common. The color bar indicates the number of APOGEE stars in the cluster (saturating at 5). The square symbols denote clusters with a single star in \citet{zhang_19}.}
 	\label{fig:lamost}
 \end{figure}

\citet{zhang_19} published mean abundances for open clusters using results from the LAMOST survey \citep{lamostdr1}. Our sample includes 22 open clusters in common with \citet{zhang_19} and we find a median offset in [Fe/H] (in the sense LAMOST - APOGEE) of -0.01  {dex;} however we note some significant outliers. Figure \ref{fig:lamost} shows the difference in [Fe/H] between \citet{zhang_19} and this work ([Fe/H]$_{LAMOST}$ - [Fe/H]$_{APOGEE}$). There is fairly good agreement near solar metallicity, but towards lower metallicities (as measured by APOGEE), there are some clusters with highly discrepant results.
 {The three clusters with the most discrepant metallicities, $\gtrsim 0.2$ dex, are Czernik 23, ASCC 21, and NGC 2264 (in increasing order by their APOGEE [Fe/H]). The two clusters off by $\sim 0.4$ dex (Czernik 23 and NGC 2264) have only one star in the \citet{zhang_19} analysis, and Czernik 23 has only one star in APOGEE as well. NGC 2264 and ASCC 21 are among the young clusters which were excluded from our high quality sample.}
Removing these  {three} most discrepant clusters, the LAMOST values are much more consistent with APOGEE. 

A previous comparison of APOGEE DR14 to LAMOST found an offset in [Fe/H] of 0.06 with a scatter of 0.13 \citep{anguiano_2018}. Given the analysis in \S\ref{sec:dr14comp}, it is not surprising that APOGEE DR16 appears to be in better agreement with LAMOST. 

%%%%%%%%%%%%%%%%%%%%%%%%%%%%%%%%%%%%%%%%%%%%%%%%%%%%%%%%%%%%%%%%%%%%%%%%%%%%%%%%%%%%%%%%%%%%%%%%%
%%%%%%%%%%%%%%%%%%%%%%%%%%%%%%%%%%%%%%%%%%%%%%%%%%%%%%%%%%%%%%%%%%%%%%%%%%%%%%%%%%%%%%%%%%%%%%%%%
%\clearpage
\section{Measuring Galactic Trends}

\subsection{Choosing a Distance Catalog}\label{sec:distProb}

In OCCAMII, Galactocentric distances to open clusters were calculated using the average distance for member stars from \citet{bailer_jones_cat}. However, due to the application of a geometric prior to each star individually, this may not be an optimal solution for clusters \citep{bailer_jones_cat}. Another uniform source of distances is therefore desired. 

\begin{figure}
    \epsscale{1.1}
 	\plotone{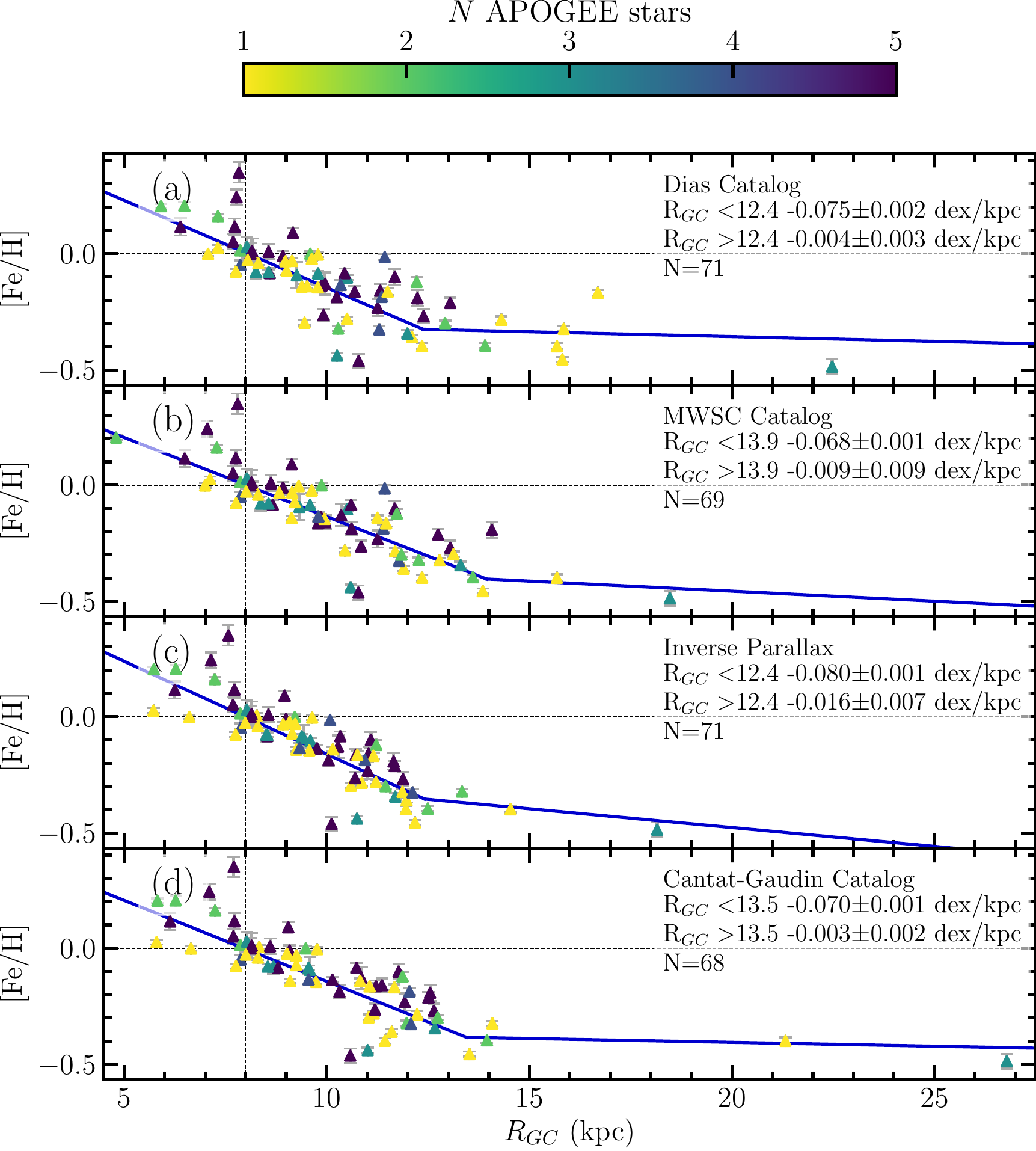}
 	%\vskip-0.35in
 	\caption{ \small [Fe/H] vs \rgc trends measured using different distance determinations. This is similar to an analysis performed in OCCAMII, but we have added measurements from \citet{cg_18} where available. The colorbar shows the number of APOGEE stars per cluster, saturating at 5.}
 	\label{fig:distComp}
\end{figure}

Distances to open clusters are frequently recomputed by many groups. Some form of isochrone fitting has been used by a number of studies \citep[e.g.,][]{vonHippel_06, mwsc_catalog}, however, only \citet[MWSC]{mwsc_catalog} have produced a catalog using a uniform isochrone fitting method to measure distances for a very large (over 1000) set of open clusters. Recently, the \gaia survey has made it possible to create large catalogs of cluster distances based on parallax \citep[e.g.,][]{cg_18}. Of the two large catalogs, the MWSC catalog covers significantly more of our sample, but still,  {two clusters in our high quality sample (BH 211 and Teutsch 12)} are not included. For these clusters, we rely on stellar parallaxes from {\em Gaia} DR2. Since the MWSC catalog does not include distance uncertainties, we assume an uncertainty of 10\% of the distance. 

For completeness, and to highlight the significant influence that choosing a particular  {distance} catalog can have on the measured gradient, Figure \ref{fig:distComp} repeats the basic analysis of \S \ref{sec:feh_grad} using three other distance catalogs (the catalogs of \citet{dias_catalog} and \citet{cg_18}, as well as inverse parallax as discussed in OCCAMII). The difference in the measured gradients is much less severe than in OCCAMII (where it was $\sim$40\%), but it is still significant, potentially as large  as $\sim$15\%. 

\subsection{Fitting Galactic Abundance Gradients}\label{sec:fitting}

It has become common in the literature, when measuring Galactic metallicity gradients, to divide the sample somewhere between \rgc $\approx 10$ kpc and \rgc $\approx 13$ kpc and fit two separate lines to the data \citep[e.g.,][]{twarog_1997, sestito2008, friel2010, jacobson_2011, carrera_2011, yong_2012, frinchaboy_13, reddy_16, magrini_2017}, with a much shallower trend in the outer galaxy than in the inner galaxy. Since the OCCAM sample includes open clusters as far away as $R_{GC} \approx 19$ kpc, we can investigate if the Galactic metallicity gradient becomes significantly shallower at a given $R_{GC}$.  

In this study, we fit two separate lines to the data, and impose the additional constraint that both must meet at some ``knee'', although the location of the knee is allowed to vary. If we let $k$ be the $x$-coordinate of the knee, the equation describing the fit line is then:

\begin{equation} \label{line}
y = \begin{cases} 
      m_1 \cdot x + b_1 & x \leq k \\
      m_2 \cdot (x-k) + (m_1 \cdot k + b_1) & x > k 
     \end{cases}
\end{equation}

We estimate the values of $m_1$, $b_1$, $m_2$, and $k$ using maximum likelihood estimation. Uncertainties in each parameter are estimated using the \textit{emcee} package \citep{emcee}. For trends which do not appear to have multiple components (e.g., [$\alpha$/Fe] vs $R_{GC}$ trends), we perform a maximum likelihood fit and \textit{emcee} error estimation for a  {single} line. 

\section{The Galactic Metallicity Gradient}\label{sec:feh_grad}

\begin{figure*}[t!]
 	\begin{center}
 		\vskip-0.3in
 		%\includegraphics[width=\linewidth]
%         \epsscale{1.0}
     \plotone{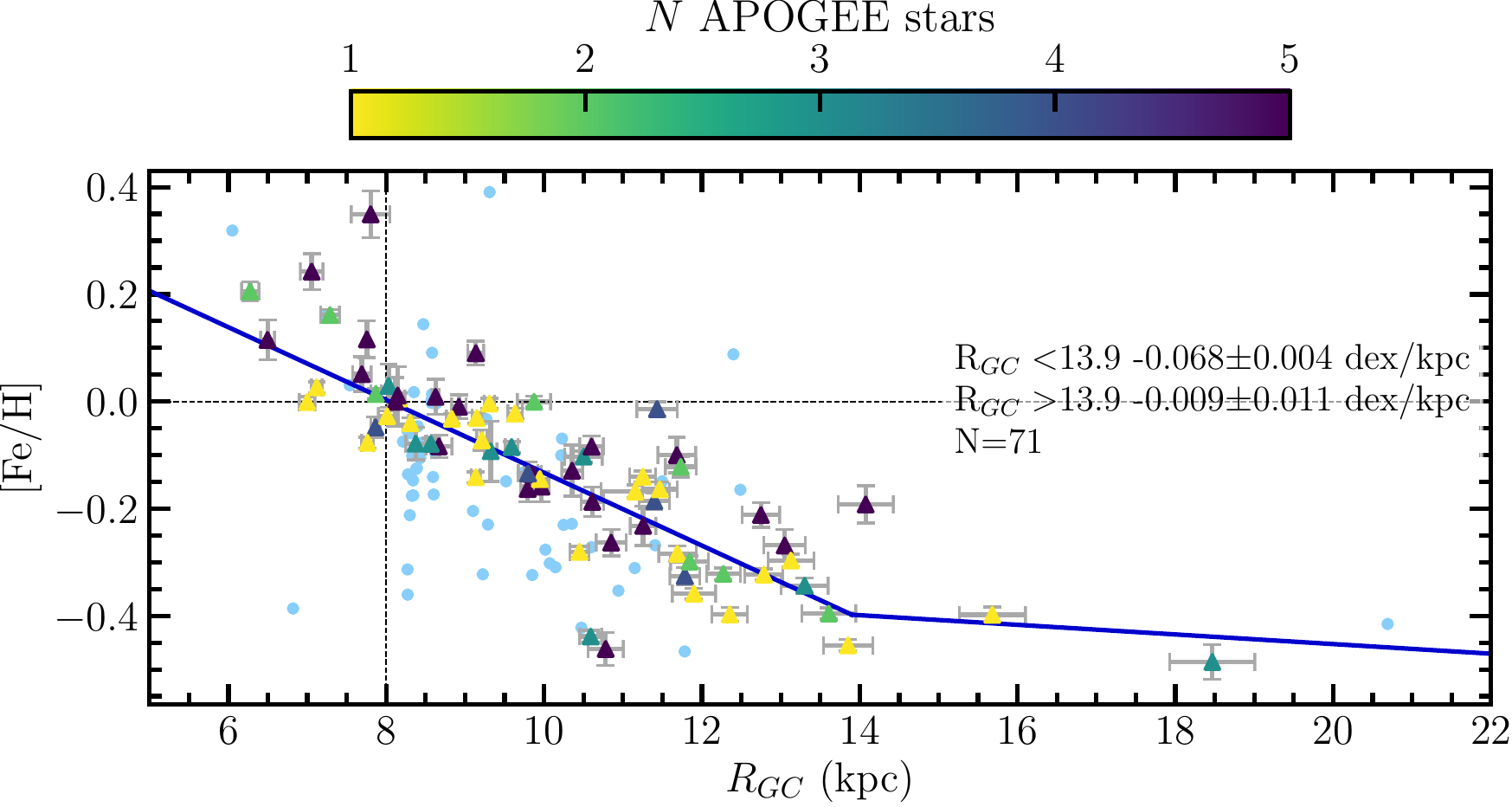}
 	\end{center}
 	%\vskip-0.35in
 	\caption{ \small The full high quality sample Galactic [Fe/H]  {versus} \rgc trend, with a 2-line fit (described by Eq. \ref{line}). Clusters flagged with quality ``0'' are shown as light blue circles. The color bar indicates the number of member stars per cluster, saturating at 5.}
 	\label{fig:feh_grad}
 \end{figure*}

Fitting to the overall [Fe/H]  {versus} \rgc gradient using open clusters as probes is common in many Galactic studies ( {see} e.g., Table 4 of OCCAMII). 
We fit the overall [Fe/H] vs \rgc trend using our high quality sample of  {71} open clusters, with a 2 line function fit (Figure \ref{fig:feh_grad}). We find an inner (\rgc$< 13.9$ kpc) gradient of $-0.068 \pm 0.004$ dex/kpc and an outer (\rgc$> 13.9$ kpc) gradient of $-0.009 \pm 0.011$ dex/kpc.

A consensus on the apparent location of the ``knee'' has nearly been reached in the literature, with values converging around \rgc $\approx 12$ kpc. However, this location does not appear to have been rigorously tested {;} that is, the position of the ``knee'' has never been included as a free parameter in the fit.

We find the location of the break in the Galactic [Fe/H] vs \rgc trend to be at \rgc $= 13.9$ kpc. To our knowledge
this is the first study to fit the ``knee'' as a free parameter.  {However, as shown in Figure \ref{fig:distComp}, this is dependent on the distance catalog adopted, and we recognize the poor coverage of our sample in the region \rgc $> 14$ kpc and the effect this may have on the determination of this parameter. }
 {Additional} open clusters in this \rgc range have been targeted as part of APOGEE 2 and should be observed soon.

%%%%%%%%%%%%%%%%%%%%%%%%%%%%%%%%%%%%%%%%%%%%%%%%%%%%%%%%%%%%%%%%%%%%%%%%%%%%%%%%%%%%%%%%%%%%%%%%%
%%%%%%%%%%%%%%%%%%%%%%%%%%%%%%%%%%%%%%%%%%%%%%%%%%%%%%%%%%%%%%%%%%%%%%%%%%%%%%%%%%%%%%%%%%%%%%%%%

If we consider only the 19 open clusters studied in OCCAM II and fit a single line as in that previous study, we find a gradient of $-0.047 \pm 0.005$ dex/kpc if we include NGC 6791 and $-0.041 \pm 0.005$ dex/kpc if we  {do} not include NGC 6791. OCCAM II found a gradient of $-0.044 \pm 0.003$ dex/kpc using distances from the MWSC catalog and excluding NGC 6791.
We emphasize that although we find a global offset of 0.05 dex in [Fe/H] between DR14 and DR16, this is not expected to have an effect on the slope of the [Fe/H]  {versus} \rgc trend as the offset should be roughly similar at any given [Fe/H]. Given the comparison between gradients derived from DR14 and DR16 results, this appears to be the case.

Table 4 of OCCAMII summarized recent measurements of the Galactic metallicity gradient from the literature in the distance range considered of $6 \lesssim$ \rgc $\lesssim 14$ kpc,  {and revealed} a range of gradients between $-0.052$ dex/kpc to $-0.085$ dex/kpc. The result in this study of $-0.068$ dex/kpc  {sits} neatly in the middle of this range. 

We can compare in more detail to the recent results from  {\citet{carrera_18}}, which also used APOGEE data (from DR14). The authors chose to split their sample at \rgc $=11$ kpc, and find an inner gradient of $-0.077 \pm 0.007$ dex/kpc. This is nearly in agreement with our result. We note the authors used distances from  {\citet{cg_18}}; in Figure \ref{fig:distComp}d we measure the metallacity gradient using the same distances and find a slope of $-0.070 \pm 0.001$ dex/kpc, in good agreement with their result.

%%%%%%%%%%%%%%%%%%%%%%%%%%%%%%%%%%%%%%%%%%%%%%%%%%%%%%%%%%%%%%%%%%%%%%%%%%%%%%%%%%%%%%%%%%%%%%%%%
%%%%%%%%%%%%%%%%%%%%%%%%%%%%%%%%%%%%%%%%%%%%%%%%%%%%%%%%%%%%%%%%%%%%%%%%%%%%%%%%%%%%%%%%%%%%%%%%%

\section{Galactic Trends for Other Elements}

\subsection{Galactic Trends for $\alpha$-Elements}

\begin{figure}
    \epsscale{1.2}
 	\plotone{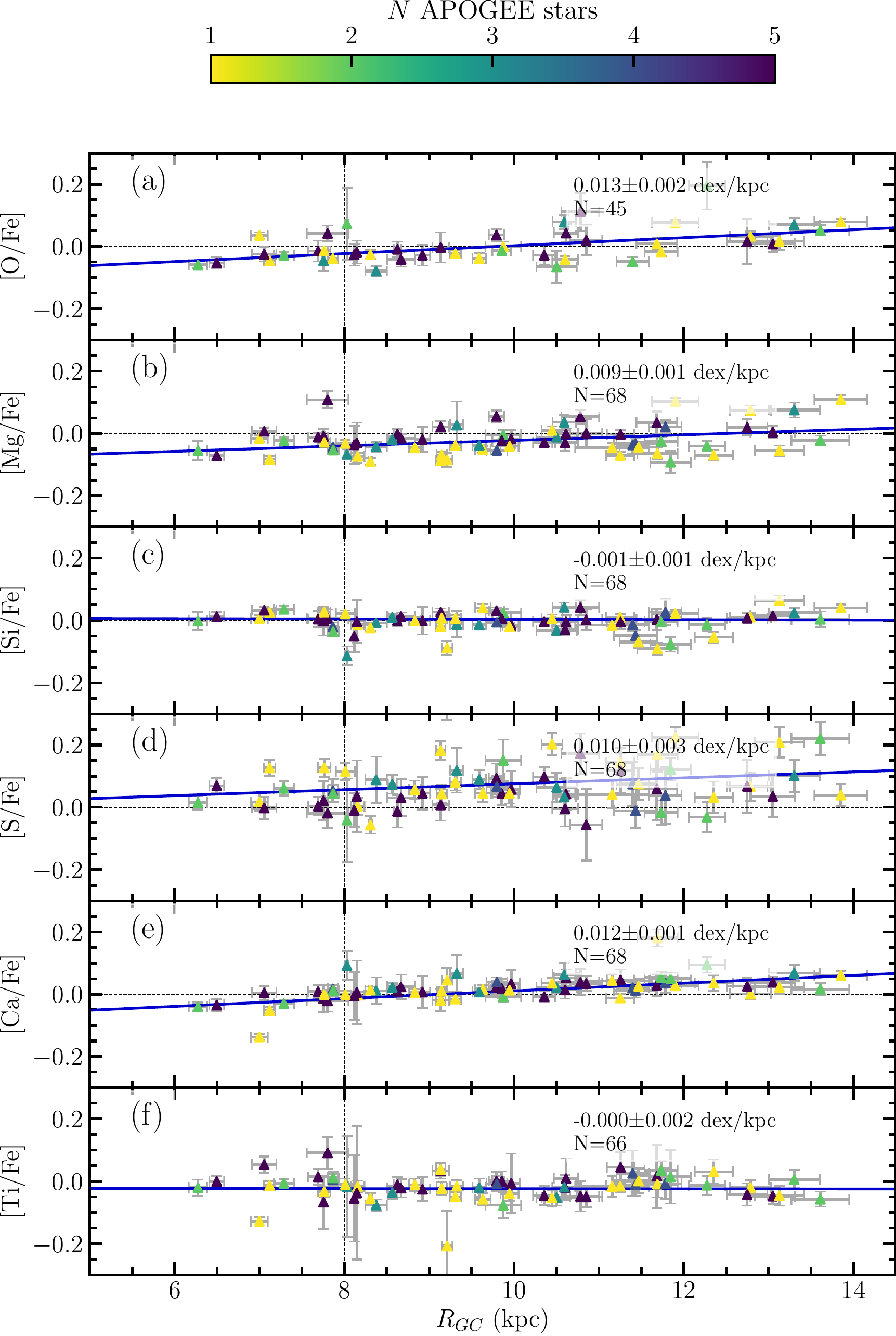}
 	%\vskip-0.35in
 	\caption{ \small The [X/Fe] vs \rgc trend for $\alpha$ elements. As before the color bar indicates number of member stars, saturating at 5.}
 	\label{fig:alpha}
 \end{figure}

Figure \ref{fig:alpha} shows Galactic trends  {versus Fe} for six $\alpha$-elements (O, Mg, S, Si, Ca, and Ti). Since we find a break in the [Fe/H] vs \rgc trend at \rgc $\approx 14$ kpc, we  {limit} the sample  {to} \rgc $<14$ kpc and measure trends for the inner clusters. 
For all $\alpha$ elements studied here, except for silicon and titanium, there is a statistically significant slight positive trend from the inner galaxy to the outer galaxy and the gradients in [$\alpha$/Fe] are consistent overall. However, for silicon and titanium we find a flat gradient. We note there is significant scatter for [S/Fe], and very little scatter for [Ca/Fe]. 

Our results are consistent with \citet{yong_2012} who measured mild positive gradients for [O/Fe], [Si/Fe], and [Ca/Fe], of the order of 0.01 dex kpc$^{-1}$,  {but} a flat trend for [Mg/Fe], although the uncertainties on all four trends are nearly as large as their measured gradients. \citet{casamiquela_2019} report slight positive gradients for [Si/Fe] ($0.022 \pm 0.007$) and [Mg/Fe] ($0.011 \pm 0.01$) in their uniform sample of open clusters in the range $6 \le R_{GC} \le 11$ kpc, although both slopes are much shallower when they include more clusters from the literature. \citet{carrera_2011} and \citet{reddy_16} report [$\alpha$/Fe] vs \rgc gradients of $0.004 \pm 0.001$ dex kpc$^{-1}$ and $0.014 \pm 0.005$ dex kpc$^{-1}$, respectively. Our results are therefore in good agreement with the literature, except perhaps for Si which appears to be almost completely flat in our case.

 {Recent work using APOGEE data showed a possible temperature effect for silicon abundances \citep{zasowski_2019}: cooler stars show lower abundances than warmer ones. The stars in more distant clusters tend to be cooler since only brighter, more evolved stars are detectable farther away. Thus the flat [Si/Fe] trend may partly reflect this effect in APOGEE data.}

%%%%%%%%%%%%%%%%%%%%%%%%%%%%%%%%%%%%%%%%%%%%%%%%%%%%%%%%%%%%%%%%%%%%%%%%%%%%%%%%%%%%%%%%%%%%%%%%%
%%%%%%%%%%%%%%%%%%%%%%%%%%%%%%%%%%%%%%%%%%%%%%%%%%%%%%%%%%%%%%%%%%%%%%%%%%%%%%%%%%%%%%%%%%%%%%%%%

\subsection{Galactic Trends for Iron-Peak Elements}\label{sec:ironPeak}

 \begin{figure}
    \epsscale{1.2}
 	\plotone{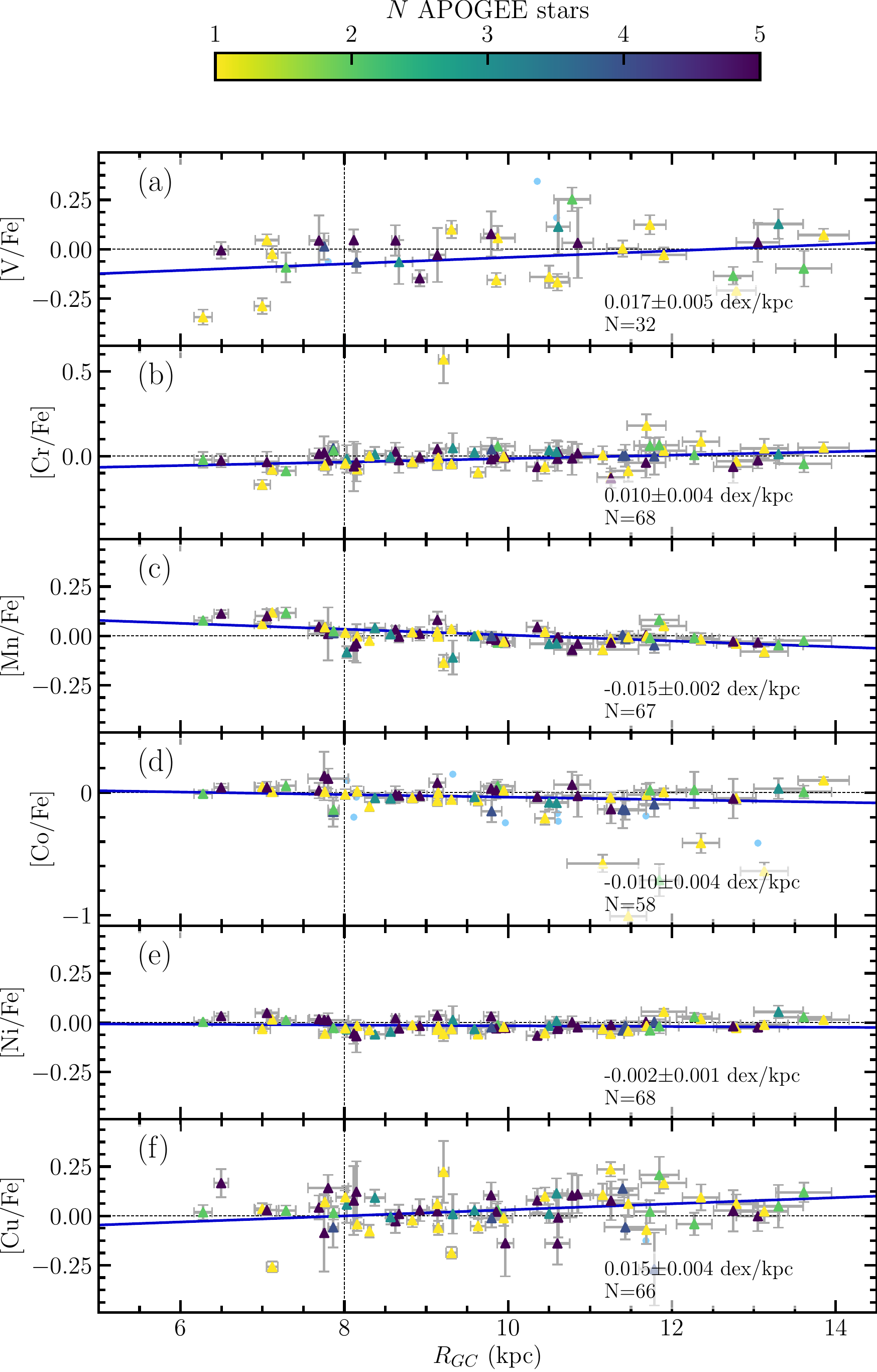}
 	%\vskip-0.35in
 	\caption{ \small The [X/Fe] vs \rgc trend for iron-peak elements. Light blue circles are clusters that have an [X/Fe] abundance reported but $\sigma$ [X/Fe] $\ge 0.2$ dex.}
 	\label{fig:iron_peak}
 \end{figure}

APOGEE DR16 reports abundances for six elements that are classified as ``iron-peak'' elements: vanadium, chromium, manganese, cobalt, nickel, and copper. Figure \ref{fig:iron_peak} shows Galactic abundance trends for each of these elements.
 {The [Ni/Fe] vs \rgc trend is completely flat;} the abundances stay very near solar with small scatter for the  {Galactic radii} explored.
Statistically significant slightly positive trends are measured for [V/Fe], [Cr/Fe], and [Cu/Fe], however there are some significant outliers for [Cr/Fe] (Czernik 18 having a single star with [Cr/Fe] $=+0.57$) and [Cu/Fe] (Chupina 1 having a single star with [Cu/Fe] $=-0.58$). 
There is a statistically significant, slightly negative trend measured for [Co/Fe], however a number of outliers to this trend are present between $R_{GC} \approx 11$ kpc and $R_{GC} \approx 13$ kpc.
Interestingly, \citet{casamiquela_2019} find a mildly significant \textit{negative} trend for [V/Fe]. For [Cr/Fe] they find conflicting trends depending on which sample they use. This suggests a need for more observational  {data to better constrain the} gradients in these elements. 

For [Mn/Fe], a significant negative trend of -0.015 $\pm$ 0.002 is found. We note this is consistent with the trend first presented in OCCAMII. \citet{yong_2012} find a [Mn/Fe] gradient of -0.06 $\pm$ 0.01 in the region $R_{GC} < 13$ kpc, but this measurement is made using only $\sim 8$ open clusters. 
Since this trend is not well studied, little discussion  {of it} exists in the literature.
A relatively simple explanation may be that higher [Mn/Fe] abundances in the inner Galaxy are the result of larger contributions to chemical enrichment from type Ia supernovae (SNe Ia) \citep{nomoto_2013}, perhaps suggesting less recent star formation towards the inner galaxy or higher SNe Ia efficiency in the inner Galaxy.

%%%%%%%%%%%%%%%%%%%%%%%%%%%%%%%%%%%%%%%%%%%%%%%%%%%%%%%%%%%%%%%%%%%%%%%%%%%%%%%%%%%%%%%%%%%%%%%%%
%%%%%%%%%%%%%%%%%%%%%%%%%%%%%%%%%%%%%%%%%%%%%%%%%%%%%%%%%%%%%%%%%%%%%%%%%%%%%%%%%%%%%%%%%%%%%%%%%

\subsection{ {``Odd-z''} Gradients}

\begin{figure}
    \epsscale{1.1}
 	\plotone{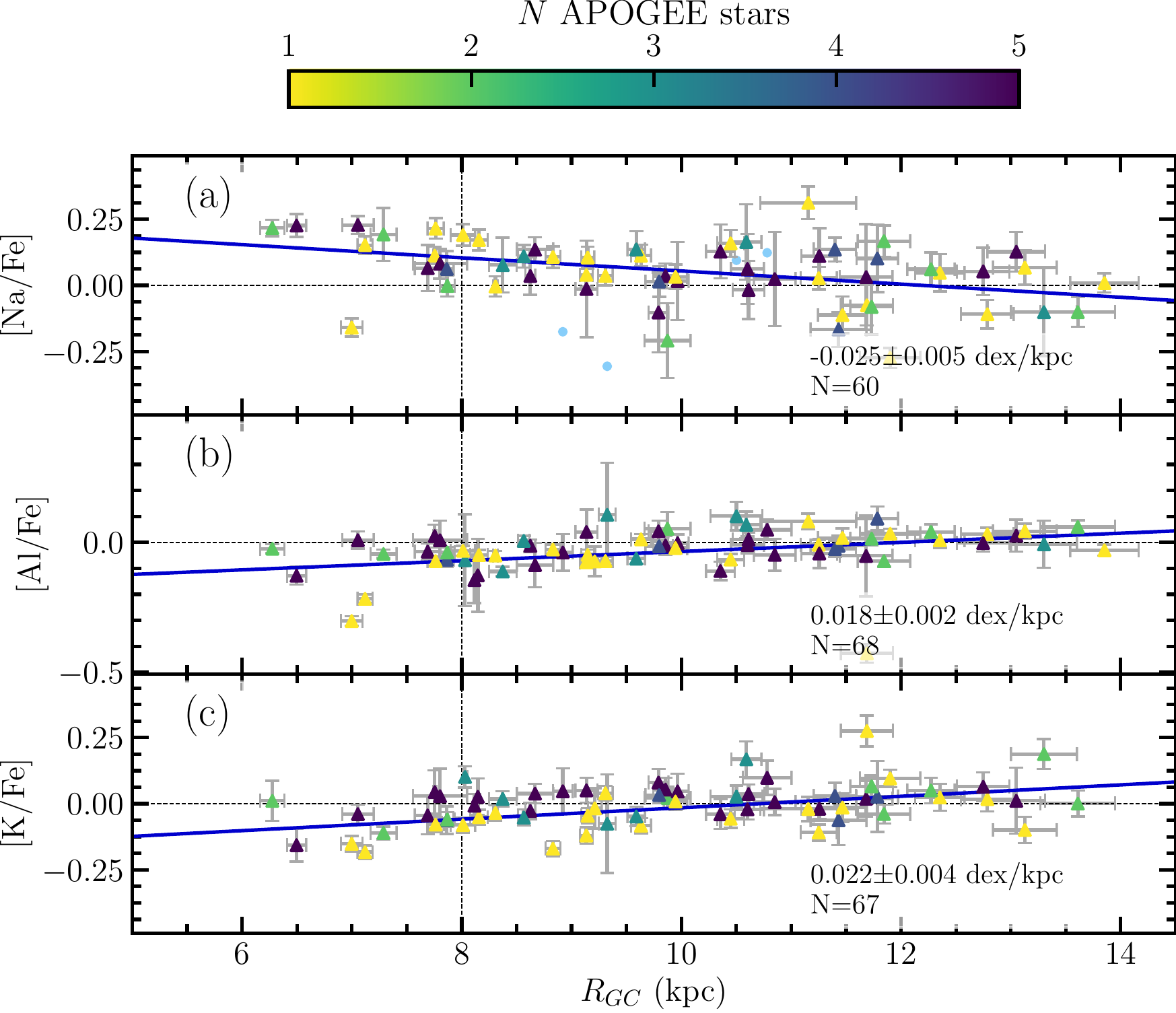}
 	%\vskip-0.35in
 	\caption{ \small The [X/Fe] vs \rgc trend for the  {"odd-z"} elements reported in APOGEE DR16. As before, the color bar indicates number of members and light blue circles are clusters with very high uncertainty in that element.}
 	\label{fig:random}
 \end{figure}

There are three other  {APOGEE} elements that  {do not} readily fall into the above categories: sodium, aluminum, and potassium {, often referred to as ``odd-z'' elements}. We note that while [P/Fe] abundances are reported in DR16, there are serious doubts about the reliability of the abundances for this element (see J\"onsson et al. \textit{in prep}). Figure \ref{fig:random} shows the Galactic trends for [Na/Fe], [Al/Fe],  {and} [K/Fe]. 
[Al/Fe] and [K/Fe] show nearly identical significant positive gradients, while
[Na/Fe] shows a significant negative gradient.  All three trends have at least one significant outlier, but the trends nevertheless appear fairly  {robust}.
\citet{yong_2012} find a similar trend for [Al/Fe] of 0.03 $\pm$ 0.01 dex/kpc; for [Na/Fe], however, they find a flat trend with significant scatter.

\section{The Evolution of Galactic Abundance Gradients}

\citet{minchev_19} discuss the effect that sample selection can have on measured  {abundance} gradients,  {in particular the bias introduced by most samples containing a majority of young clusters.}
To more accurately compare to previous work, and provide more meaningful comparisons for galactic evolution models,  {in this section} we compare mono-age samples.

\subsection{Iron}\label{sec:fe_evo}

\begin{figure}
    \epsscale{1.1}
 	\plotone{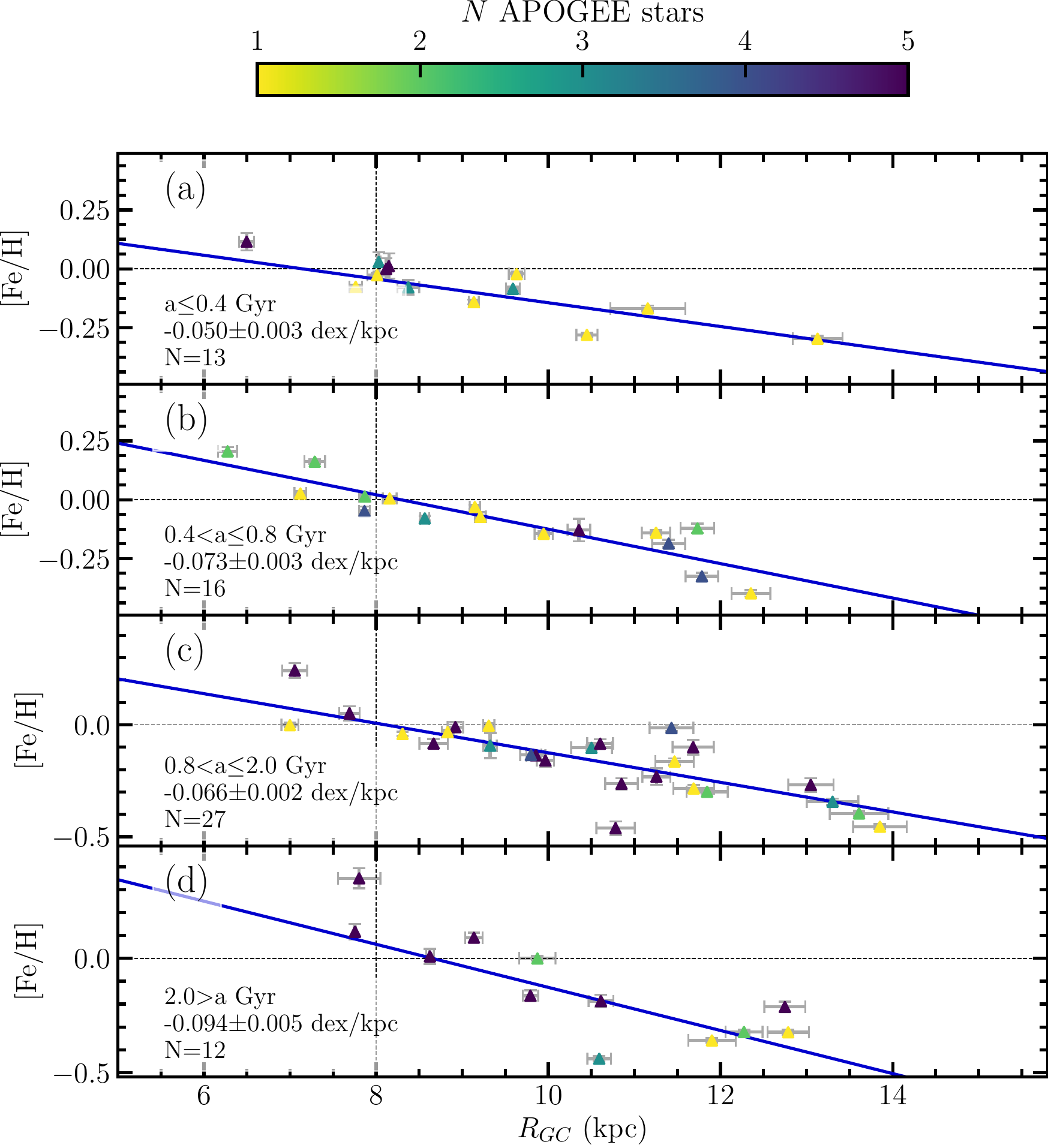}
 	%\vskip-0.35in
 	\caption{ \small The Galactic [Fe/H] vs \rgc trend in 4 age bins, showing the general decrease in steepness over time.}
 	\label{fig:age}
\end{figure}

Our sample is large enough that it can be split into four age bins,  {which we divide} at 400 Myr, 800 Myr, and 2 Gyr, with all bins being reasonably well populated.  
Figure \ref{fig:age} shows the [Fe/H]  {versus} \rgc trend for clusters  {separated} in age bins. We use ages from the MWSC catalog  {because} they are derived in a uniform fashion, and should certainly be reliable enough to place clusters in the coarse bins we have chosen. 

\begin{figure}
    \epsscale{1.3}
 	\plotone{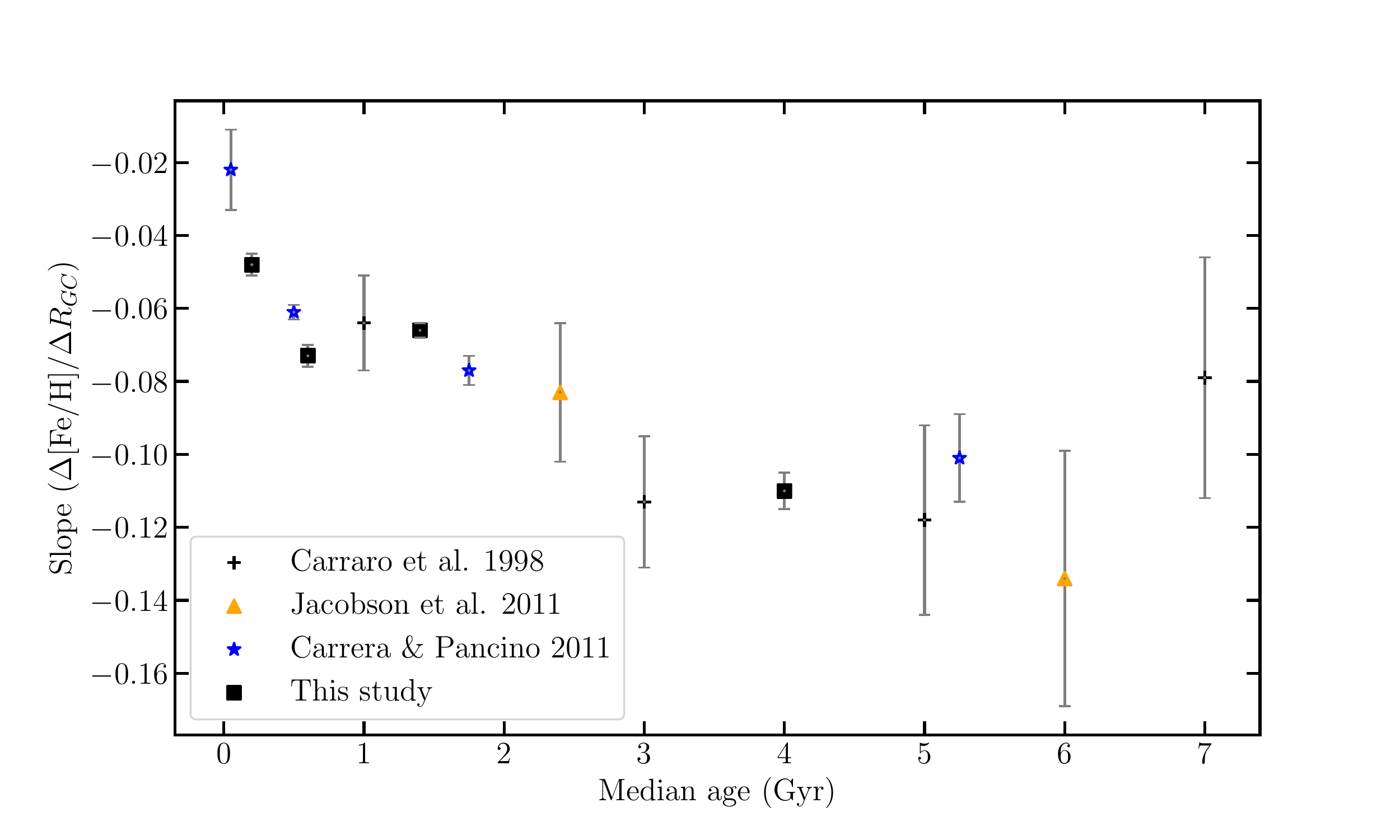}
 	%\vskip-0.35in
 	\caption{ \small A summary of Galactic metallicity gradients measured in mono-age populations from the literature.}
 	\label{fig:lit_comp_age}
\end{figure}

The evolution of the [Fe/H] vs \rgc trend has been studied extensively in the literature \citep[e.g.][]{carraro_98, friel_02, jacobson_2011, carrera_2011, yong_2012}. A summary of results from the literature is provided in Figure \ref{fig:lit_comp_age}. Here we plot the measured metallicity gradient for clusters in a given age range vs the middle of that age range (for example the middles of our age bins are 0.2, 0.6, 1.4, 4 Gyr). It is important to note that the majority of clusters from all four studies in Figure \ref{fig:lit_comp_age} fall in the range \rgc $< 14$ kpc, with the exception of a few clusters from \citet{carraro_98}.
Figure \ref{fig:lit_comp_age} shows a consistent trend of steeper metallicity gradients for older populations. There is one point in disagreement with this trend: the oldest clusters from \citet{carraro_98} appear to reverse this trend. This may be due to the inclusion of some clusters near $R_{GC} \approx 15$ in their oldest bin. If we consider the large uncertainties on the two oldest measurements, it is possible the trend levels out after 4 Gyr.

It should be mentioned that the trend found here is opposite  {that} seen for field stars \citep[e.g.,][]{anders_2017}, where the oldest populations show a shallower gradient. Radial migration is expected to cause this flattening of the metallicity gradient on a long enough time scale \citep[e.g.,][]{minchev_2018}. To explain the absence of this phenomenon in open clusters \citet{anders_2017} suggest that clusters that do not migrate or clusters that migrate towards the inner Galaxy preferentially break up.

\begin{figure*}
    \epsscale{1.2}
 	\plotone{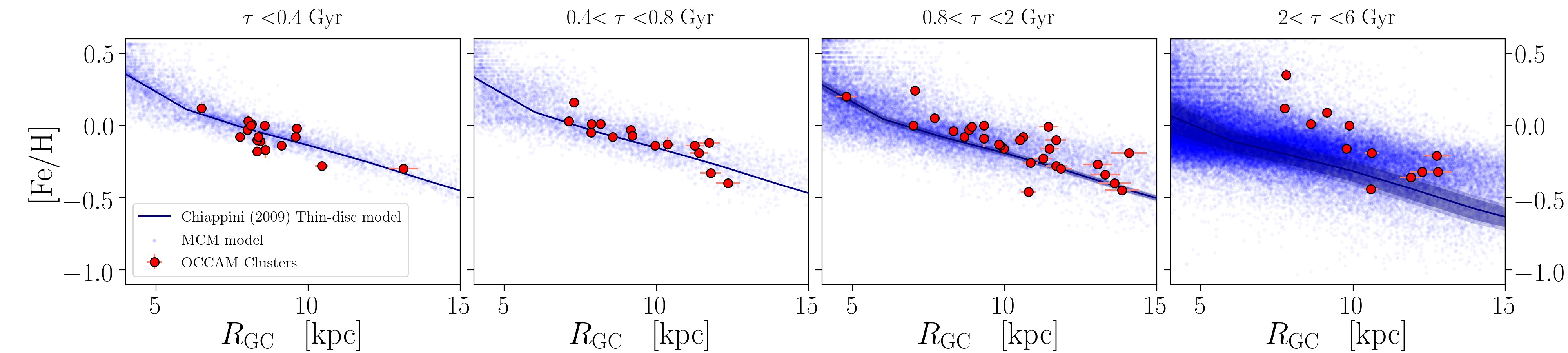}
 	%\vskip-0.35in
 	\caption{OCCAM IV clusters (red) plotted with the pure chemical evolution model of \citet{chiappini_2009} (blue line) and the MCM chemo-dynamical simulation  {\citep{minchev_1, minchev_2}}, seperated into the age bins used previously.}
 	\label{fig:models}
\end{figure*}

In Figure \ref{fig:models} we show the OCCAM IV sample plotted with the pure chemical evolution model of \citet{chiappini_2009} and the chemo-dynamical simulation of \citet[MCM]{minchev_1, minchev_2}, divided in the same age bins as Figure \ref{fig:age}.
 {There is good agreement between the models and the OCCAM IV sample in the younger three bins.}
In the oldest bin the effects of radial migration are clearly seen in the MCM points. Also in the oldest bin, there is a noticeable lack of clusters towards the inner galaxy,  {and} a clear steepening of the gradient,  {which could be} due to migration of inner old clusters towards outer regions. This is consistent with the suggestion from \citet{anders_2017} that clusters migrating inward preferentially break up. Elsewhere, the clusters are roughly consistent with the MCM model.

\subsection{Other Elements}

\begin{figure*}[t!]
    \begin{center}
    \epsscale{1.2}
 	\plotone{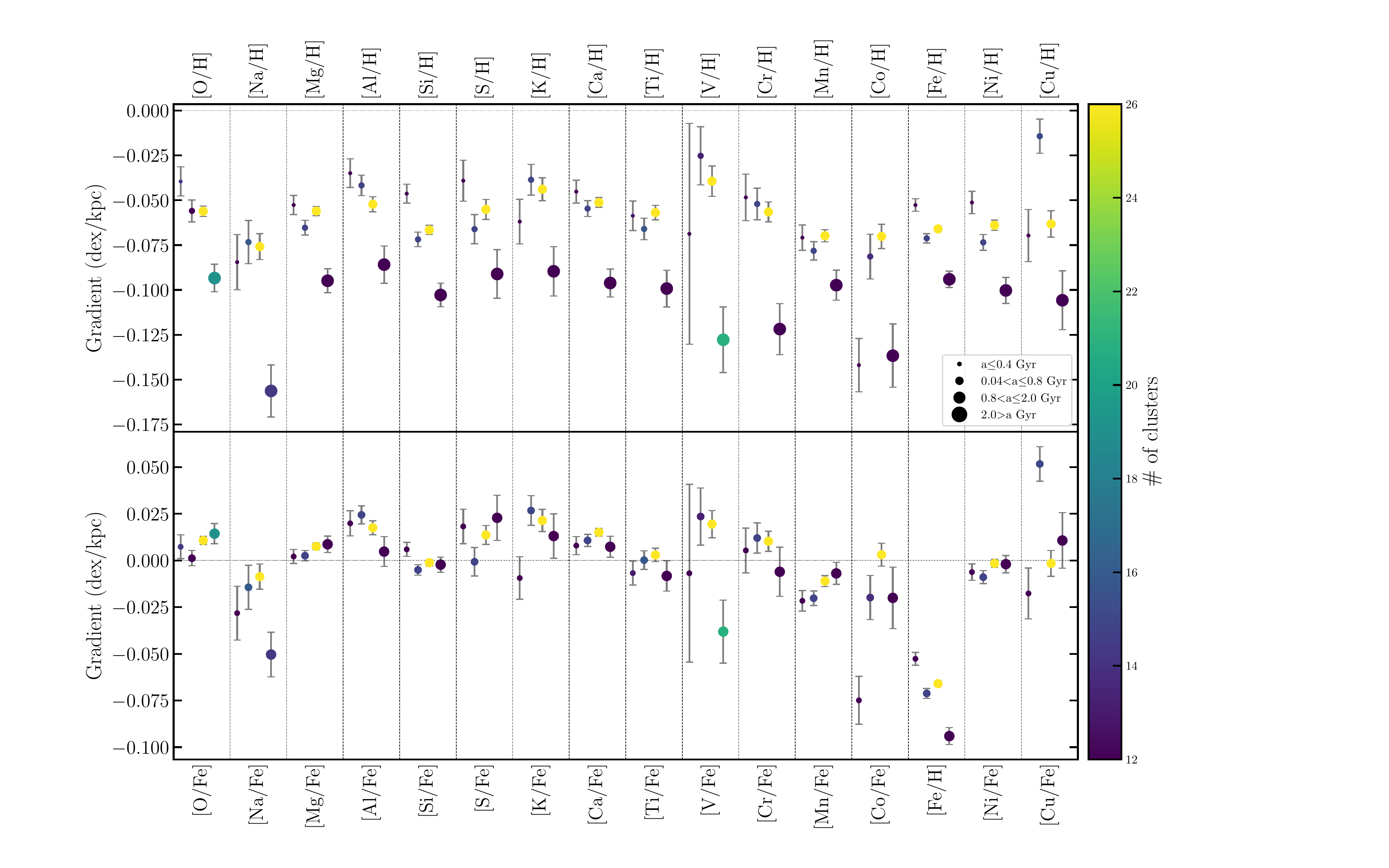}
 	\vskip0.1in
 	\caption{ \small Gradients measured in four age bins as for Figure \ref{fig:age} are plotted for each element. The points increase in size from youngest to oldest;  {the color indicates number of clusters used to measure each gradient}.}
 	\label{fig:ageSum}
 	\end{center}
\end{figure*}

Age trends in elements  {other than} iron also provide insight into the chemical evolution of the Galaxy. The top panel of Figure \ref{fig:ageSum} provides a summary of abundance gradients for each element presented previously as a function of cluster age, measured in the same four age bins as for iron (Figure \ref{fig:age}). 
The top panel of Figure \ref{fig:ageSum} shows an overall similar behavior for all elements: the gradient for the oldest population (open clusters older than $\sim 2$ Gyr) is the steepest; this is reminiscent of what was observed for [Fe/H]. For [Na/H], [Ti/H], [Cr/H], and [Mn/H] we cannot distinguish between the gradients measured for the intermediate-age and young populations. For [O/H], [Mg/H], [Si/H], [S/H], [K/H], [V/H], [Co/H], [Fe/H], and [Ni/H] the youngest population shows a distinct {ly} flatter gradient, but the two intermediate-age populations are indistinguishable within the uncertainties. Relatively flat $\alpha$-element abundance gradients have also been found for young B stars \citep[e.g.,][]{daflon_2004} and H II regions \citep[e.g.,][]{Esteban_15}.
We note  {that} for [V/H] the youngest bin is populated with only  {five} clusters,  {while} for [Co/H] the gradient in the youngest population is heavily influenced by a single very [Co/H]-poor cluster. 

\subsection{The Evolution of [X/Fe] Gradients}

To understand  {the} differences in the evolution of elemental abundances  {better}, it is also informative to study the evolution of [X/Fe] gradients over time. The bottom panel of figure \ref{fig:ageSum} is similar to the top panel but now we show the evolution of [X/Fe] trends. A variety of trends can be seen; some elements show a stable trend over time (e.g., [Ni/Fe], [Si/Fe]), some show an increasingly positive trend (e.g., [Al/Fe]), and [Mn/Fe] shows an increasingly negative trend. All of these trends are worth discussing and we do so below. We do not consider [V/Fe], [Cr/Fe], [Co/Fe], or [Cu/Fe] in detail, either because the uncertainties are larger than the trends or because one or more age bins are poorly populated for that element. We do not discuss [Ni/Fe] further because, as stated  {in \S\ref{sec:ironPeak}}, Ni appears to track Fe closely.

The results in Figure \ref{fig:ageSum} show no clear evidence of evolution in the [$\alpha$/Fe] gradients within the time spanned by this open cluster sample.
It could be argued that there are mild trends with age for the [O/Fe], [Mg/Fe], and [Ca/Fe] gradients, but the changes between different aged populations are on the order of the uncertainties. For [Si/Fe], [S/Fe], and [Ti/Fe] there are more variations, but also larger uncertainties and it is more appropriate to consider the gradients  {as} roughly constant for different aged populations. It was shown in Figure \ref{fig:alpha} that nearly all of the [$\alpha$/Fe] abundances exhibit mildly increasing radial trends and Figure \ref{fig:ageSum} indicates that such trends appear to be fairly stable within the time spanned by  {our cluster} sample. 
The flattening of the abundance gradients in recent times suggests more recent chemical enrichment in the outer Galaxy, but, taken together with the stability of the increasing [$\alpha$/Fe] gradient, we might deduce that the enrichment in the outer Galaxy had a more significant contribution from core-collapse supernovae. 
This is consistent with the conclusions from \S\ref{sec:ironPeak} and  {the discussion} below, that supernovae Ia dominated recent enrichment in the inner Galaxy.

For [Na/Fe] and [Al/Fe], the gradients for the oldest clusters are clearly set apart, even considering the sizeable uncertainty. For [Al/Fe], in particular, there appears to be a clear trend where we see the younger populations showing an increasingly positive slope.
Significantly larger Na and Al yields are expected from core collapse supernovae than SNe Ia \citep{nomoto_2013}, so a flattening of the [Na/Fe] gradient and an increasingly positive [Al/Fe] gradient are both consistent with either more recent star formation in the outer Galaxy than the inner Galaxy or higher SNe Ia efficiency in the inner Galaxy. This is also consistent with the explanation for the [Mn/Fe] gradient in \S\ref{sec:ironPeak}.

Figure \ref{fig:ageSum} shows that the [Mn/Fe] gradient becomes more negative for younger cluster populations. \citet{yamaguchi_15} showed that SNe Ia yields of manganese are strongly dependent on progenitor metallicity; higher metallicity progenitors will yield significantly more manganese. So as metals build up in the inner Galaxy, a higher [Mn/Fe] abundance is expected. This may explain the evolution of the [Mn/Fe] gradient in general terms.

%%%%%%%%%%%%%%%%%%%%%%%%%%%%%%%%%%%%%%%%%%%%%%%%%%%%%%%%%%%%%%%%%%%%%%%%%%%%%%%%%%%%%%%%%%%%%%%%%
%%%%%%%%%%%%%%%%%%%%%%%%%%%%%%%%%%%%%%%%%%%%%%%%%%%%%%%%%%%%%%%%%%%%%%%%%%%%%%%%%%%%%%%%%%%%%%%%%

\section{Conclusions }

We present a sample of 128 open clusters,  {71} of which we designate ``high quality'', using APOGEE DR16. We demonstrate that DR16  {cluster abundances are} in good agreement with  {those of} other high resolution abundance studies. Using the high quality sample, we measure Galactic abundance gradients in  {16 chemical} elements, and we measure how those gradients change for different age samples.

We find an overall Galactic [Fe/H] vs \rgc gradient of $-0.068 \pm 0.004$ dex kpc$^{-1}$ for \rgc $< 13.9$ kpc, but we re-emphasize the point of OCCAMII that this result can vary significantly depending on which catalog of distances is used. We show general agreement with the literature in regards to the evolution of this gradient. 

For the first time, we fit the knee in the Galactic abundance gradient  {as a free parameter} at $R_{GC} = 13.9$ kpc, but we recognize a need for more clusters beyond this break to more reliably constrain the fit.

We find general agreement with the literature for gradients in $\alpha$ elements. 
We present further evidence for the negative [Mn/Fe] vs \rgc trend first found in OCCAMII. 
We find significant Galactic trends in  {vanadium,} chromium, and copper, although we are unable to suggest a strong explanation for these trends.  
We find very significant trends in sodium, aluminum, and potassium; so-called ``odd-Z'' elements. We recognize a need for further study of trends in these elements as they are not well reported in the literature.

We divide our sample into four age bins and investigate changes in  {16} elements over time. We show that  {[X/H]} abundance gradients for all  {16} elements follow the same general trend, becoming more shallow over time, as has consistently been found for iron. We further investigate age trends in [X/Fe] for  {15} elements. A number of these trends seem to support a similar conclusion: either increased SNe Ia efficiency towards the inner Galaxy or less recent star formation in the inner Galaxy compared to the outer Galaxy.

\acknowledgements

We would like to thank Marina Kounkel for very helpful discussions about young clusters. We would also like to thank Jos\'e G. Fern\'andez-Trincado and Borja Anguiano for helpful comments.

JD and PMF acknowledge support for this research from the National Science Foundation (AST-1311835 \& AST-1715662).
KC acknowledges support for this research from the
National Science Foundation (AST-0907873).

DAGH acknowledges support from the State Research Agency (AEI) of the Spanish Ministry of Science, Innovation and Universities (MCIU) and the European Regional Development Fund (FEDER) under grant AYA2017-88254-P.

D.G. and D.M. gratefully acknowledge support from the Chilean Centro de Excelencia en Astrof\'isica
y Tecnolog\'ias Afines (CATA) BASAL grant AFB-170002.
D.G. also acknowledges financial support from the Dirección de Investigaci\'on y Desarrollo de
la Universidad de La Serena through the Programa de Incentivo a la Investigaci\'on de
Académicos (PIA-DIDULS).
D.M. is also supported by the Programa Iniciativa Cientifica Milenio grant IC120009, awarded to the Millennium Institute of Astrophysics (MAS), and by Proyecto FONDECYT regular No. 1170121.

H. J. acknowledges support from the Crafoord Foundation, Stiftelsen Olle Engkvist Byggm\"astare, and Ruth och Nils-Erik Stenb\"acks stiftelse.

A. Roman-Lopes acknowledges financial support provided in Chile by Comisi\'on Nacional de Investigaci\'on Cient\'ifica y Tecnol\'ogica (CONICYT) through the FONDECYT project 1170476 and by the QUIMAL project 130001

Funding for SDSS-III has been provided by the Alfred P. Sloan Foundation, the Participating Institutions, the National Science Foundation, and the U.S. Department of Energy Office of Science. The SDSS-III web site is http://www.sdss3.org/.

SDSS-III is managed by the Astrophysical Research Consortium for the Participating Institutions of the SDSS-III Collaboration including the University of Arizona, the Brazilian Participation Group, Brookhaven National Laboratory, Carnegie Mellon University, University of Florida, the French Participation Group, the German Participation Group, Harvard University, the Instituto de Astrofisica de Canarias, the Michigan State/Notre Dame/JINA Participation Group, Johns Hopkins University, Lawrence Berkeley National Laboratory, Max Planck Institute for Astrophysics, Max Planck Institute for Extraterrestrial Physics, New Mexico State University, New York University, Ohio State University, Pennsylvania State University, University of Portsmouth, Princeton University, the Spanish Participation Group, University of Tokyo, University of Utah, Vanderbilt University, University of Virginia, University of Washington, and Yale University.

Funding for the Sloan Digital Sky Survey IV has been provided by the Alfred P. Sloan Foundation, the U.S. Department of Energy Office of Science, and the Participating Institutions. SDSS-IV acknowledges
support and resources from the Center for High-Performance Computing at
the University of Utah. The SDSS web site is www.sdss.org.

SDSS-IV is managed by the Astrophysical Research Consortium for the 
Participating Institutions of the SDSS Collaboration including the 
Brazilian Participation Group, the Carnegie Institution for Science, 
Carnegie Mellon University, the Chilean Participation Group, the French Participation Group, Harvard-Smithsonian Center for Astrophysics, 
Instituto de Astrof\'isica de Canarias, The Johns Hopkins University, 
Kavli Institute for the Physics and Mathematics of the Universe (IPMU) / 
University of Tokyo, Lawrence Berkeley National Laboratory, 
Leibniz Institut f\"ur Astrophysik Potsdam (AIP),  
Max-Planck-Institut f\"ur Astronomie (MPIA Heidelberg), 
Max-Planck-Institut f\"ur Astrophysik (MPA Garching), 
Max-Planck-Institut f\"ur Extraterrestrische Physik (MPE), 
National Astronomical Observatories of China, New Mexico State University, 
New York University, University of Notre Dame, 
Observat\'ario Nacional / MCTI, The Ohio State University, 
Pennsylvania State University, Shanghai Astronomical Observatory, 
United Kingdom Participation Group,
Universidad Nacional Aut\'onoma de M\'exico, University of Arizona, 
University of Colorado Boulder, University of Oxford, University of Portsmouth, 
University of Utah, University of Virginia, University of Washington, University of Wisconsin, 
Vanderbilt University, and Yale University.

This work has made use of data from the European Space Agency (ESA) mission {\it Gaia} (\url{https://www.cosmos.esa.int/gaia}), processed by the {\it Gaia} Data Processing and Analysis Consortium (DPAC, \url{https://www.cosmos.esa.int/web/gaia/dpac/consortium}). Funding for the DPAC has been provided by national institutions, in particular the institutions participating in the {\it Gaia} Multilateral Agreement.

This research made use of Astropy, a community-developed core Python package for Astronomy (Astropy Collaboration, 2018).

\vspace{5mm}
\facilities{Sloan (APOGEE), FLWO:2MASS, \gaia}

\software{\href{http://www.astropy.org/}{Astropy}}, 

\bibliography{Donor.bib}

\clearpage

\end{document}